\documentclass[a4paper,11pt]{article}

\usepackage{jheppub}
\usepackage[utf8]{inputenc}
\usepackage{comment}
\usepackage[section]{placeins}
\usepackage{amsmath}
\usepackage{multirow,makecell}
\usepackage{tabularx}
\usepackage{booktabs}
\usepackage{graphicx}
\usepackage{braket}
\usepackage{color}
\usepackage{float}
\usepackage{chemformula}
\usepackage{siunitx}
\usepackage{url}
\usepackage{xspace}
\usepackage{caption}
\usepackage{soul}

\newcommand\mbb{m_{\beta\beta}}

\newcommand\onbb{$0\nu\beta\beta$\xspace}

\newcommand\psf{G}

\newcommand\MeV{$\si{\mega\electronvolt}$\xspace}

\newcommand\meV{$\si{\milli\electronvolt}$\xspace}

\newcommand\xe{$\ch{^{136}Xe}$}
\newcommand\te{$\ch{^{130}Te}$}
\newcommand\ger{$\ch{^{76}Ge}$}
\newcommand\mo{$\ch{^{100}Mo}$}
\newcommand\se{$\ch{^{82}Se}$}

\title{Impact of nuclear matrix element calculations for current and future neutrinoless double beta decay searches \\
}

\author[a]{Federica Pompa,}
\author[b]{Thomas Schwetz,}
\author[b,c]{and Jing-Yu Zhu}
\affiliation[a]{Instituto de F\'{i}sica Corpuscular (IFIC), University of Valencia-CSIC, Parc Cient\'{i}fic UV, c/Catedr\'{a}tico Jos\'{e} Beltr\'{a}n 2, E-46980 Paterna, Spain}
\affiliation[b]{Institut für Astroteilchen Physik, Karlsruher Institut für Technologie (KIT), Hermann-von-Helmholtz-Platz 1, 76344 Eggenstein-Leopoldshafen, Germany
}
\affiliation[c]{School of Physics and Astronomy and Tsung-Dao Lee Institute, Shanghai Jiao Tong University, 800 Dongchuan Rd, Shanghai 200240, China}
\emailAdd{fpompa@ific.uv.es, schwetz@kit.edu, zhujingyu@sjtu.edu.cn}
\abstract{Nuclear matrix elements (NME) are a crucial input for the interpretation of neutrinoless double beta decay data. We consider a representative set of recent NME calculations from different methods and investigate the impact on the present bound on the effective Majorana mass $m_{\beta\beta}$ by performing a combined analysis of the available data as well as on the sensitivity reach of future projects. A crucial role is played by the recently discovered short-range contribution to the NME, induced by light Majorana neutrino masses. Depending on the NME model and the relative sign of the long- and short-range contributions, the current $3\sigma$ bound can change between $m_{\beta\beta} < 40$~meV and 600~meV. The sign-uncertainty may either boost the sensitivity of next-generation experiments beyond the region for $\mbb$ predicted for inverted mass ordering or prevent even advanced setups to reach this region. 
Furthermore, we study the possibility to distinguish between different NME calculations by assuming a positive signal and by combining measurements from different isotopes. Such a discrimination will be impossible if the relative sign of the long- and short-range contribution remains unknown, but can become feasible if $m_{\beta\beta} \gtrsim 40$~meV and if the relative sign is known to be positive. Sensitivities will be dominated by the advanced $^{76}$Ge and $^{136}$Xe setups assumed here, but NME model-discrimination improves if data from a third isotope is added, e.g., from $^{130}$Te or $^{100}$Mo.}

\begin{document}
\maketitle
\section{Introduction}
One of the most fundamental questions in particle physics is whether neutrinos are Dirac or Majorana particles. If they were Majorana particles, we might observe neutrinoless double beta decay (\onbb), the lepton-number violating decay of a nucleus with mass number $A$ and charge number $Z$: 
\begin{equation}
    \label{eq:AZ}
    (A,Z) \longrightarrow (A,Z+2) + 2 e^-\,.
\end{equation}
Currently, this is the only known feasible way to prove the Majorana property of neutrinos and a lot of experimental as well as theoretical efforts have been put towards its discovery, see e.g.\ Refs.~\cite{Dolinski:2019nrj,Agostini:2022zub} for recent reviews. Apart from its fundamental role to test whether lepton number is conserved or not, this process provides also information on the absolute size of the neutrino mass, complementary to alternative probes from nuclear beta decay \cite{KATRIN:2019yun,KATRIN:2021uub}, cosmology \cite{DiValentino:2021hoh,Planck:2018vyg}, and maybe a future supernova observation \cite{Hansen:2019giq,Pitik:2022jjh,Pompa:2022cxc}.
Here, we only consider the simplest mechanism of Eq.~\eqref{eq:AZ} due to light neutrino exchange. One can refer to Refs. \cite{Fang:2021jfv,Dekens:2023iyc} for different effects involving sterile neutrinos in neutrino mass models.

However, the relation between the \onbb\ decay rate and the neutrino mass is subject to large theoretical uncertainties, as it requires the knowledge of nuclear matrix elements (NME). 
With $\Gamma_\alpha$ and $(T_{1/2})_\alpha$ standing, respectively, for the \onbb decay rate and its half life of a given nucleus labeled by the index $\alpha$, we have
\begin{equation}
\label{eq:decay_rate}
(T_{1/2}^{-1})_\alpha =\widetilde\Gamma_\alpha(\mbb,M_{\alpha i})=
\frac{\Gamma_\alpha(\mbb,M_{\alpha i})}{\ln 2} = G_\alpha \, |M_{\alpha i}|^2 \, m_{\beta\beta}^2 \;,
\end{equation}
where we also introduce $\widetilde\Gamma_\alpha(\mbb,M_{\alpha i})$ as the inverse half life of \onbb for later calculation, $\psf_\alpha$ is commonly known as the Phase-Space Factor (PSF) \cite{Kotila_2012, Deppisch:2020ztt} and encloses the kinematics of the decay, $M_{\alpha i}$ denotes the NME where the index $i$ labels different nuclear models for the NME calculations, and $\mbb$ is the effective Majorana mass, which is related to fundamental neutrino properties by
\begin{equation}
    \label{eq:mbb}
    \mbb = \left| \sum_j U_{ej}^2 m_j \right| \,.
\end{equation}
The sum runs over light neutrino mass states (light compared to the typical energy scale of the process, which is of order \MeV), with $U_{ej}$ being an element of the Pontecorvo–Maki–
Nakagawa–Sakata (PMNS) mixing matrix and $m_j$ are the Majorana masses of the neutrino mass states. In particular, if the three-neutrino mass spectrum has the so-called inverted mass ordering (IMO), neutrino oscillation data predicts a minimal allowed range for $\mbb$, by assuming that the lightest neutrino mass is zero:
\begin{equation}
    \label{eq:mbb-IMO}
    14 \,{\rm meV} < \mbb < 49\,{\rm meV} \qquad\text{(minimal IMO range)} \,,
\end{equation}
where the interval emerges due to the unknown Majorana phases in $U_{ej}$ and includes the $3\sigma$ range for the oscillation parameters \cite{Esteban:2020cvm}. The $\mbb$ range in Eq.~\eqref{eq:mbb-IMO} provides a specific benchmark goal for the next generation of \onbb\ projects.

It is clear from Eq.~\eqref{eq:decay_rate}, that information on $\mbb$ can only be extracted from a measurement of $\Gamma_\alpha$ if the NME is known. Different techniques are applied for the calculation of NME and the corresponding results differ by factors of a few, see Refs. \cite{Agostini:2022zub,ejiri_neutrino-mass_2020, Engel:2016xgb} for reviews. Recently it has been noticed \cite{Cirigliano:2018hja,Cirigliano:2019vdj}, that in addition to the known long-range contribution of light neutrino masses to the NME, also a short-range contribution to the NME appears at leading order in chiral-perturbation theory. Its presence is required for a consistent renormalization of the amplitude. 
This new contribution is not related to heavy lepton-number violating beyond-standard model physics inducing \onbb\ decay (see e.g. \cite{Deppisch:2020ztt}) but appears already in the minimal scenario, with only light Majorana neutrino masses being present. 
In our work we consider only the light Majorana neutrino exchange mechanism, while new short-range contributions may also appear in beyond-standard model scenarios, for example the heavy neutrino exchange mechanism, e.g.\  \cite{Agostini:2022zub,Dekens:2023iyc}. 
This new short-range contribution introduces an additional uncertainty on the value of the NME. Apart from its numerical value, also the relative sign of long- and short-range contributions is not known, which can lead either to positive or negative interference. This can have substantial impact on the interpretation of experimental results in terms of neutrino masses \cite{Jokiniemi:2021qqv}.

In this paper we first consider present constraints on \onbb\ 
obtained by the GERDA \cite{GERDA:2020xhi} and MAJORANA \cite{PhysRevLett.130.062501} collaborations based on $\ch{^{76}Ge}$, 
the CUORE collaboration \cite{CUORE:2021mvw} from $\ch{^{130}Te}$ and by the KamLAND-Zen~\cite{KamLAND-Zen:2022tow} and EXO-200~\cite{PhysRevLett.123.161802} collaborations from $\ch{^{136}Xe}$. For a given model for the NME, the results from different isotopes can be combined to provide a joint constraint on $\mbb$. We will study the dependence of the combined constraint on the different NME models and highlight the impact of the short-range contribution. 

Then we study several planned future \onbb\ projects and investigate their discovery reach for $\mbb$ as a function of NME models, with and without taking into account the short-range contribution. Assuming that future projects can establish a positive signal for \onbb, the combination of measurements from different isotopes, in principle may allow to discriminate between NME models. We study this question quantitatively and investigate which NME models could be excluded or established by \onbb\ observations of next-generation of experiments.

The outline of the paper is as follows. In Sec.~\ref{sec:NME} we give a brief overview of the different NME calculations and the recently discovered short-range contribution due to light Majorana neutrino masses. In Sec.~\ref{sec:current} we present the global analysis of current \onbb\ results and study the impact of NME models on the combined upper bound on $\mbb$. In Sec.~\ref{sec:future} we introduce a set of advanced next-generation experiments and investigate their sensitivity to $\mbb$ as a function of the different NME models, whereas in Sec.~\ref{sec:NMEdiscr} we address the question whether a combination of several future experiments using different isotopes can experimentally distinguish between different NME calculations, assuming a positive signal has been observed. We conclude in Sec.~\ref{sec:conclusions}. The appendix provides supplementary information on the NME discrimination power of different combinations of next-generation experiments. Through out this paper we assume that light Majorana neutrino masses are the only mechanism responsible for \onbb\ decay.

\section{Nuclear models and nuclear matrix elements} \label{sec:NME}

Uncertainties on \onbb NME calculations derive mainly from the nuclear theory side. 
In fact, even when \onbb can be observed someday, NMEs can only be obtained via numerical methods \cite{ejiri_neutrino-mass_2020, Engel:2016xgb}.
Different groups performing refined many-body simulations based on the Nuclear Shell Model (NSM) \cite{Menendez:2017fdf,Horoi:2015tkc,Coraggio:2020hwx,Coraggio:2022vgy}, the Quasiparticle Random-Phase Approximation (QRPA) \cite{Mustonen:2013zu,Hyvarinen:2015bda,Simkovic:2018hiq,Fang:2018tui,Terasaki:2020ndc}, the Energy-Density Functional theory (EDF) \cite{Rodriguez:2010mn,LopezVaquero:2013yji,Song:2017ktj} and the Interacting Boson Model (IBM) \cite{Barea:2015kwa,Deppisch:2020ztt} provide, for each relevant isotope, the NME value according to the respective model.

Refs.~\cite{Cirigliano:2018hja,Cirigliano:2019vdj} pointed out a new contribution to NME, required to obtain a consistent renormalization of the \onbb\ amplitude due to light-neutrino exchange. A contribution due to short-range interaction (SRI) has to appear at leading order to cancel divergences of the standard long-range part, see also \cite{Jokiniemi:2021qqv,Cirigliano:2020dmx,Cirigliano:2021qko}. We parameterize the total NME for a given nucleus $\alpha$ and the specific nuclear model calculation $i$ as
\begin{equation}
    \label{eq:M_short_long}
   M_{\alpha i} = M_{\alpha i}^{\text{long}} + M_{\alpha i }^{\text{short}} = M_{\alpha i }^{\text{long}}(1 + n_{\alpha i}) \,, 
\end{equation}
where $M_{\alpha i}^{\text{long}}$ ($M_{\alpha i}^{\text{short}}$) denotes the long-range (short-range) contribution to the NME, and we have defined
\begin{equation}
\label{eq:n}
n_{\alpha i} = \frac{M_{\alpha i }^{\text{short}}}{M_{\alpha i}^{\text{long}}} \,,
\end{equation}
expressing the relative impact of the short-range contribution to \onbb.
Since both its value and sign are unknown, the correction due to the short-range term could either enhance or suppress the expected decay rate. 


\begin{table}[t]
\centering
\caption{Compilation of $M_{\alpha i}^\text{long}$ values 
for light Majorana neutrino exchange calculated with different nuclear models
from \cite{Agostini:2022zub}. 
These results have been obtained by assuming the bare value of the axial coupling constant $g_A^{\rm{free}} = 1.27$.
Each model is identified through an index, given in the second column.}
\label{tab:NME}
\begin{tabular}{ccccccc}
\toprule
\makecell[c]{Nuclear Model}  & \makecell[c]{Index [Ref.]} & $\ch{^{76}Ge}$& $\ch{^{82}Se}$ & $\ch{^{100}Mo}$ & $\ch{^{130}Te}$ & $\ch{^{136}Xe}$  \\
\midrule
\multirow{5}*{NSM} & N1 \cite{Menendez:2017fdf} & $2.89$ & $2.73$ & -& $2.76$ &  $2.28$   \\
                            &  N2 \cite{Menendez:2017fdf}  & $3.07$ & $2.90$ &-&   $2.96$ &  $2.45$ \\
                            &  N3 \cite{Horoi:2015tkc} & $3.37$  & $3.19$ & - & $1.79$ &  $1.63$\\
                            &  N4 \cite{Horoi:2015tkc} & $3.57$ & $3.39$ & -& $1.93$ &  $1.76$  \\
                            &  N5  \cite{Coraggio:2020hwx,Coraggio:2022vgy}& $2.66$ & $2.72$ & $2.24$& $3.16$ &  $2.39$ \\
\midrule
\multirow{5}*{QRPA} & Q1 \cite{Mustonen:2013zu}&$5.09$ & - & -& $1.37$ &  $1.55$  \\
                            &  Q2 \cite{Hyvarinen:2015bda} & $5.26$& $3.73$ & $3.90$  & $4.00$ &  $2.91$  \\
                            &  Q3 \cite{Simkovic:2018hiq} & $4.85$ & $4.61$ & $5.87$& $4.67$ &  $2.72$ \\
                            & Q4 \cite{Fang:2018tui} & $3.12$ & $2.86$ & - & $2.90$ &  $1.11$ \\
                            &  Q5 \cite{Fang:2018tui} & $3.40$  & $3.13$ & -& $3.22$ &  $1.18$ \\
                            &  Q6 \cite{Terasaki:2020ndc} & -  & - & -& $4.05$ &  $3.38$ \\
\midrule
\multirow{3}*{EDF} & E1 \cite{Rodriguez:2010mn} & $4.60$ & $4.22$ & $5.08$ & $5.13$ &  $4.20$  \\
                            & E2  \cite{LopezVaquero:2013yji} & $5.55$  & $4.67$ & $6.59$ & $6.41$ &  $4.77$\\
                            & E3  \cite{Song:2017ktj} & $6.04$ & $5.30$ & $6.48$& $4.89$ &  $4.24$  \\
\midrule
\multirow{2}*{IBM} & I1 \cite{Barea:2015kwa}& $5.14$ & $4.19$ & $3.84$& $3.96$ &  $3.25$  \\
                            &  I2 \cite{Deppisch:2020ztt} & $6.34$ & $5.21$ & $5.08$& $4.15$ &  $3.40$  \\
\bottomrule 
\end{tabular}
\end{table} 

We summarize the various long-range NME calculations which we are going to use in our study in Tab.~\ref{tab:NME}, following the review~\cite{Agostini:2022zub}. All models assume the free/bare value of the axial-vector coupling strength $g_A^{\rm free} = 1.27$ measured in neutron beta decay \cite{ParticleDataGroup:2022pth}. 
We parameterize the $g_{A}$ quenching by introducing a parameter $q$, defined by
\begin{equation}\label{eq:quenching}
  g_A^{\rm eff} = q \, g_A^{\rm free} \,,  
\end{equation}
which may modify the NMEs by a factor $q^2$ and the decay rate by $q^4$, 
leading to another uncertainty in the interpretation in terms of $\mbb$. We assume that the quenching parameter is equal for different isotopes, but depends on the NME model considered.
The deviation of $g_A^{\rm eff}$ from the free-nucleon value $g_A^{\rm free}$ appears due to nuclear many-body effects and nuclear-medium effects \cite{Suhonen:2017krv}. 
For example, for the NSM, the quenching factor $q\sim 0.7-0.8$ is needed to reconcile related theory and experiments \cite{Agostini:2022zub}. 
Given a fixed decay rate of \onbb, a smaller $q$ will obviously lead to a larger $m_{\beta\beta}$. In the following numerical calculation, we will consider the product of $m_{\beta\beta}$ and $ q^2$
to include the quenching uncertainty.

The variation in $M_\alpha^\text{long}$ shown in Tab.~\ref{tab:NME} of about a factor three highlights the nuclear theory uncertainties. With few exceptions among the considered isotopes, the NMEs obtained by different techniques follow a common trend: NSM models tend to give the smallest NMEs, while EDF the largest, with the IBM and QRPA somewhere in between.

\begin{table}[t]
\centering
\caption{Estimated ranges for the ratio of short-range to long-range contributions to the NME, $|n_{\alpha i}|$, depending on the isotope and nuclear model assumed \cite{Jokiniemi:2021qqv}.
}
\label{tab:n_ranges}
\begin{tabular}{ccc}
\toprule
\multirow{2}*{Isotope} & \multirow{1}*{NSM} & \multirow{1}*{QRPA} \\
& \% & \% \\
\midrule
$\ch{^{76}Ge}$ & $15$--$42$ & $32$--$73$ \\
$\ch{^{82}Se}$ & $15$--$41$ & $30$--$70$ \\
$\ch{^{100}Mo}$ & - & $~49$--$108$ \\
$\ch{^{130}Te}$ & $17$--$47$ & $34$--$77$ \\
$\ch{^{136}Xe}$ & $17$--$47$ & $30$--$70$ \\
\bottomrule 
\end{tabular}
\end{table}

Concerning the contribution of SRI, currently the rough evaluations for the ratio $n_{\alpha i}$ is available only for a limited number of nuclear models: the NSM 
\cite{Menendez:2017fdf,Horoi:2015tkc,Coraggio:2020hwx} and the QRPA \cite{Mustonen:2013zu,Hyvarinen:2015bda,Simkovic:2018hiq,Fang:2018tui,Terasaki:2020ndc} ones. Depending on the isotope and on the nuclear model assumed, different ranges are estimated for this contribution, which we summarize in Tab.~\ref{tab:n_ranges}  \cite{Jokiniemi:2021qqv}. 
For both NSM and QRPA, the standard matrix element $M_{\alpha i}^\text{long}$ is larger than the new term $M_{\alpha i}^\text{short}$ (maybe except for $^{100}$Mo), but in both models the contribution of the SRI is sizable, which can considerably affect the \onbb rates expected in current and future experiments, as we will show below.

While currently there is no consensus on the sign of the SRI contribution, there are some indications that $n_{\alpha}$ is positive \cite{Wirth:2021pij,Weiss:2021rig,Cirigliano:2018hja,Agostini:2022zub}. 
In our analysis we will take a phenomenological approach and investigate the implications of both possibilities. 
According to the Fig.~10 in Ref. \cite{Agostini:2022zub}, the sign of SRI for different \onbb isotopes should be the same depending on the sign of a coupling constant $g_{\nu}^{\rm NN}$, where the unknown value of $g_\nu^{\rm NN}$ also induces part of the uncertainty on the SRI contribution and needs to be determined from lepton-number violating data (real or synthetic) \cite{Jokiniemi:2021qqv}, see \cite{Davoudi:2020gxs,Davoudi:2021noh} for lattice-QCD calculations. 
The uncertain value of $g_\nu^{\rm NN}$ introduces a correlated variation of the SRI contributions for different isotopes, which accounts for part of the ranges given in Tab.~\ref{tab:n_ranges}. In our fit we ignore this correlation and allow to vary the $|n_{\alpha i}|$ for different isotopes independently within the quoted uncertainties. 
A negative value of $g_{\nu}^{\rm NN}$ (which corresponds to positive SRI for all the isotopes) is indicated in recent Refs.~\cite{Weiss:2021rig,Cirigliano:2021qko,Wirth:2021pij,Richardson:2021xiu}. Below we consider both, negative and positive signs for the SRI contribution, in order to highlight the importance to determine this sign.

\section{Results from current \onbb experiments}
\label{sec:current}

So far, experiments agree with the null-signal hypothesis, placing lower limits on the isotopes half-life $T_{1/2}$ which translates into upper bounds on the effective Majorana mass $\mbb$ for a given NME.
Current experiments, with different detector design and techniques, place limits on the \onbb\ half-life $T_{1/2}$ of different isotopes, like $\ch{^{76}Ge}$ (by the GERDA \cite{GERDA:2020xhi} and MAJORANA \cite{PhysRevLett.130.062501} collaborations), $\ch{^{130}Te}$ (by the CUORE collaboration \cite{CUORE:2021mvw}) and $\ch{^{136}Xe}$ (by the KamLAND-Zen \cite{KamLAND-Zen:2022tow} and EXO-200 \cite{PhysRevLett.123.161802} collaborations). Here we will combine these results and investigate the dependence of the joint limit on $\mbb$ for the adopted NME models. 

\begin{figure}[t]
    \centering
\includegraphics[width=0.7\textwidth]{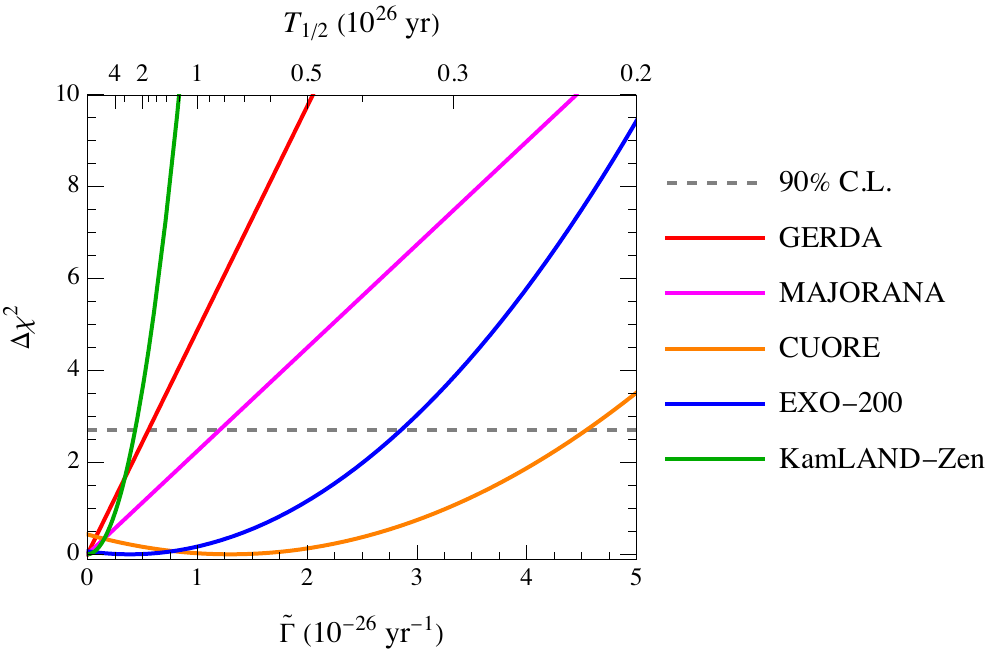}
    \caption{$\Delta\chi^2$ profiles for current experiments as a function of the \onbb inverse half time (bottom abscissa) and half-life (top abscissa) of related isotope. The dashed line indicates $\Delta\chi^2 = 2.706$, which we use as the prescription to match $90\%$~C.L.\ limits. 
    }
    \label{fig:current_lim}
\end{figure}

\begin{table}[t]
\centering
\caption{Values of the $a_r, b_r, c_r$ coefficients we obtained and used to parameterize the $\Delta\chi^2$ for each experiment according to Eq.~\eqref{eq:chi2_0nbb_curr}. 
Current $90\%$ C.L.\ lower limits on the isotope lifetime, $T_{1/2}^{90}$, in units of $10^{26}~$yr are also reported.  }
\begin{tabular}{cccccc}
\toprule
Nuclide  & Experiment & $a_r$ & $b_r$ & $c_r$ & $T_{1/2}^{90}/{10^{26}}$yr \\
\midrule
\multirow{2}*{$\ch{^{76}Ge}$}  & GERDA & $0.000$ & $4.871$ & $0.000$ & 
$1.8$ \cite{GERDA:2020xhi} \\
                         & MAJORANA & $0.000$ & $2.246$ & $0.000$ & $0.83$ \cite{PhysRevLett.130.062501} \\
      \midrule
$\ch{^{130}Te}$  & CUORE & $0.257$ & $-0.667$ & $0.433$ & $0.22$ \cite{CUORE:2021mvw} \\
\midrule
\multirow{2}*{$\ch{^{136}Xe}$}
						& KamLAND-Zen & $14.315$ & $0.000$ & $0.000$ & 
      $2.3$ \cite{KamLAND-Zen:2022tow} \\	
						& EXO-200 & $0.443$ & $-0.342$ & $0.066$ & 
      $0.35$ \cite{PhysRevLett.123.161802} \\	
\bottomrule 
\end{tabular}
\label{tab:abc_param}
\end{table} 
%
%
%
As for most of the experiments no explicit likelihood function is available, we follow the approach of
\cite{Capozzi:2021fjo,Lisi:2022nka} (see also \cite{Caldwell:2017mqu,Biller:2021bqx}) and parameterize $\Delta \chi^2 (\widetilde \Gamma_{\alpha})$ as a quadratic function:
\begin{equation}
\label{eq:chi2_0nbb_curr}
\Delta\chi^2_r(\widetilde \Gamma_{\alpha}) = a_r \,(\widetilde \Gamma_{\alpha})^2 + b_r \,\widetilde \Gamma_{\alpha} + c_r\,,
\end{equation}
where the coefficients $a_r, b_r, c_r$ can be derived for each experiment labeled by $r$, depending on the available information. This ansatz is based on the fact that the number of signal events is proportional to the inverse half life $\widetilde \Gamma_{\alpha}$. In the case of a background free experiment we expect the $\Delta\chi^2$ to depend linearly on $\widetilde \Gamma_{\alpha}$ (as it is the case for GERDA and MAJORANA), whereas in the presence of sizeable background we expect a quadratic shape.
If experiments report only upper bounds, the best fit point is assumed to be located at $\widetilde \Gamma_{\alpha} = 0$. This is the case if the upper bound coincides with the sensitivity of the experiment.\footnote{This is a good approximation e.g., for latest GERDA results. We thank B.~Schwingenheuer for private communication on this topic.}
Then the coefficients are chosen such that we can reproduce the reported 90\%~C.L.\ on  $\widetilde \Gamma_{\alpha}$ for each experiment, for which we adopt the prescription $\Delta\chi^2 = 2.706$.
Thus, based on the $90\%$ upper limit of $\widetilde \Gamma_{\alpha}$ given by the experiments, we derived their 
$\Delta\chi^2_r(\widetilde \Gamma_{\alpha})$ with the 
$a_r, ~b_r,~ c_r$ shown in Table~\ref{tab:abc_param}. More details about the derivation are as follows:
\begin{itemize}
\item For GERDA and MAJORANA, we first have $a_r =0$ and $c_r =0$ due to the linear dependence of $\Delta\chi^2_r(\widetilde \Gamma_{\alpha})$ on $\widetilde \Gamma_{\alpha}$ and best-fit point at $\widetilde \Gamma_{\alpha} =0$. We specify $T^{90}_{1/2}$ as the $90\%$ C.L.\ lower bound of the \onbb half life given by the experiment. Then $b_r$ can be obtained according to $\Delta\chi^2_r(\widetilde\Gamma^{90})= b_r \widetilde\Gamma^{90} =2.706$ with $\widetilde\Gamma^{90} = 1/T^{90}_{1/2}$.
\item In the case of CUORE, Ref.~\cite{CUORE:2021mvw} reports a best fit for $\widetilde \Gamma_{\alpha}$,
denoted by $\widetilde\Gamma^{\rm bf}$. This lead us to $\Delta\chi^2_r(\widetilde\Gamma^{90})= a_r (\widetilde\Gamma^{90}-\widetilde\Gamma^{\rm bf})^2 =2.706$ and we then obtain the coefficients $a_r, ~b_r,~ c_r$ for CUORE by solving this equation. 
\item Similarly, for EXO-200 we have the best fit of 2.7 events from the likelihood profile of the \onbb events, see Fig.~6.9 in \cite{Jewell:2020ceq}. Together with the $\widetilde\Gamma^{90}$ for EXO-200 \cite{PhysRevLett.123.161802}, we can directly get the coefficients.
\item We assume the $\widetilde\Gamma^{\rm bf} =0$ for KamLAND-Zen according to the latest results in \cite{KamLAND-Zen:2022tow} and derive the coefficients from the equation $\Delta\chi^2_r(\widetilde\Gamma^{90})= a_r (\widetilde\Gamma^{90})^2 =2.706$.
\end{itemize}
Our numbers which we report in Tab.~\ref{tab:abc_param} agree well with the ones derived in \cite{Lisi:2022nka,Capozzi:2021fjo} except for the MAJORANA and KamLAND-Zen ones, for which we updated the coefficients due to the latest results from \cite{PhysRevLett.130.062501, KamLAND-Zen:2022tow}. 
In Fig.~\ref{fig:current_lim}, we show the resulting $\Delta\chi^2$ profiles of current data, both on the inverse half-life and half-life of each isotope.

%

\begin{figure}[t]
    \centering
 \includegraphics[width=0.7\textwidth]{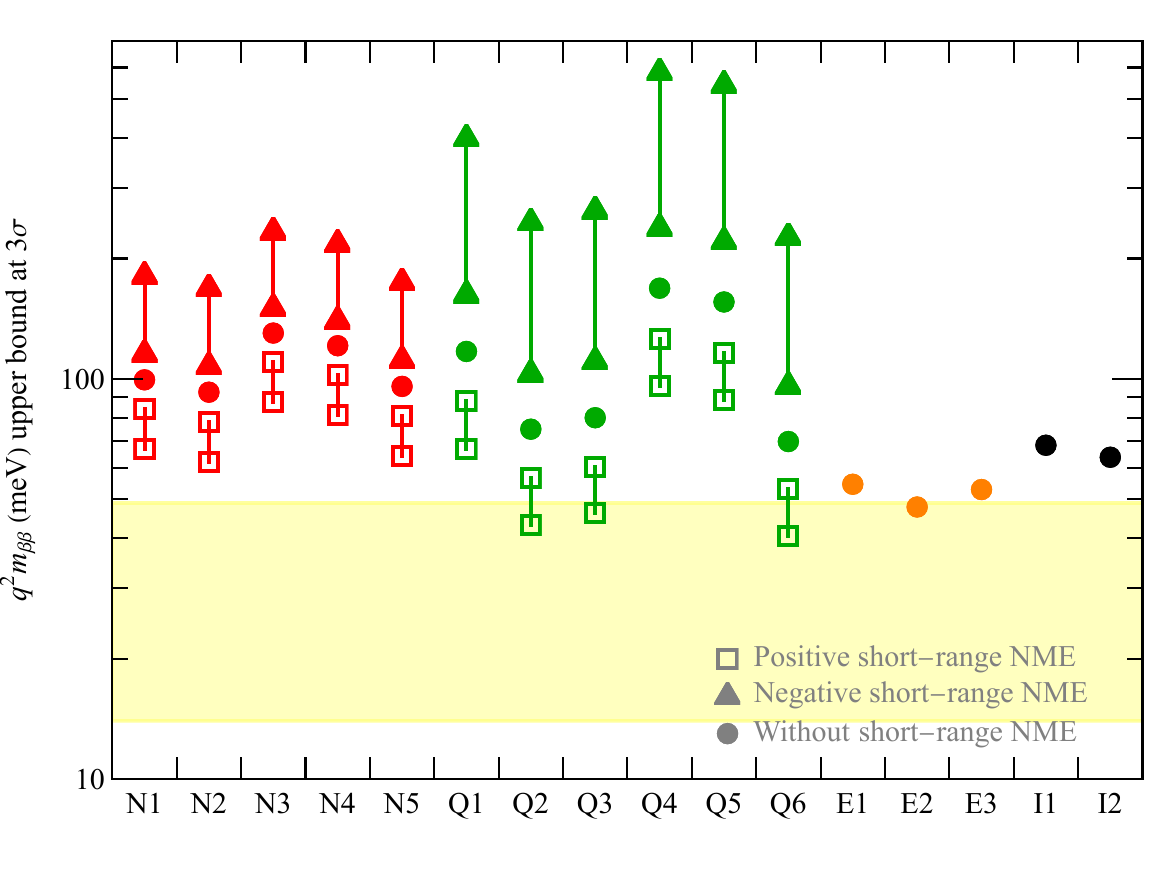}
    \caption{Upper bounds on $q^2 m_{\beta\beta} $ at $3\sigma$ C.L.\ obtained by combining current experiments listed in Tab.~\ref{tab:abc_param}, both considering and neglecting the contribution of SRI, for the NSM and QRPA models in Tab.~\ref{tab:NME}. Triangles (squares) include the contribution of SRI taken with negative (positive) sign, and vertical lines connect the upper and lower values obtained by varying $n_\alpha$ in the range of Tab.~\ref{tab:n_ranges}. For the EDF and IBM models, no short-range data are available and results considering only the long-range contribution are presented. The coloured area indicates the IMO region for $q^2 = 1$.
    }
    \label{fig:mbb_upper}
\end{figure}

Given a value for the NME, the $g_A$ quenching $q$, and the PSF, we can use Eq.~\eqref{eq:decay_rate} to translate 
$\Delta\chi^2_r(\widetilde \Gamma_{\alpha})$ into $\Delta\chi^2_r(\mbb)$, which then can be combined by summing the individual contributions:
\begin{align}
&\chi^2_\text{tot}(\mbb) = \sum_r \Delta\chi^2_r(\mbb) \,,\\
&\Delta\chi^2_\text{tot}(\mbb) = \chi^2_\text{tot}(\mbb) - \chi^2_\text{tot,min} \,. 
\end{align}

Fig.~\ref{fig:mbb_upper} shows the $3\sigma$ upper bound on $q^2\mbb$ from the combined current experiments, determined by requiring $\Delta\chi^2_{\rm tot}(\mbb)=9$. 
Here, the impact of the short-range contribution is investigated.
We consider NME calculations from the NSM and QRPA, showing the upper bounds got by ignoring (dots) or considering (triangles/squares) the contribution coming from SRI, taken with negative/positive relative sign.
Symbols represent the boundary values, while the vertical lines connect the upper and lower bounds got by varying $|n_\alpha|$ in the whole range of Tab.~\ref{tab:n_ranges}.
Strongest bounds, obtained for EDF models and QRPA with positive SRI contribution, touch already the IMO band, whereas the weakest bounds emerge from QRPA models with negative SRI.  
Note that by combining the data from various isotopes we are making use of the complementary dependence on the NMEs such that for each NME calculation the optimal bound is obtained, see \cite{Biller:2021bqx,Lisi:2022nka} for recent discussions.

Fig.~\ref{fig:mbb_upper} highlights also the rather dramatic effect which the short-range contribution to the NME can have on the upper bound on $\mbb$ (note the logarithmic scale), which can vary by up to a factor 10 in some cases. Note that, for each of the NME calculations among N1--N5 and Q1--Q6, the $q^2 m_{\beta\beta}$ limits between the two squares and between the two triangles indicated by the solid lines is obtained by adopting a fixed SRI sign for all isotopes. 

The $g_A$ quenching factor $q$ enters as multiplicative factor as $q^2\mbb $; it is therefore actually this product which is constrained by data. In order to translate the results shown in Fig.~\ref{fig:mbb_upper} into a bound on $\mbb$ alone, a value for $q$ needs to be assumed. 
Note that for $q^2$ different from one, the relative location of the upper bounds and the IMO band would be shifted accordingly; a similar comment applies to all the following figures where IMO and/or NMO bands are shown.

%
\section{Next generation of $0\nu\beta\beta$ experiments}
\label{sec:future}

\subsection{Description of experimental setups and analysis}
\label{sec:setups}

Let us now move to the discussion of future \onbb\ projects. There is a rich landscape of experiments proposed or in preparation exploiting different $\beta\beta$ emitters and experimental techniques. 
In particular, we consider LEGEND-1000 \cite{LEGEND:2021bnm} for {\ger}, SNO+II  \cite{SNO:2021xpa} for {\te}, nEXO \cite{nEXO:2021ujk} for \xe{} , SuperNEMO \cite{Kauer:2008em} for \se{} and CUPID \cite{CUPID:2022opf} for \mo{}.
Each of them is characterized by a set of so called performance parameters, through which both the expected signal and background rates can be expressed \cite{Agostini:2022zub}.
These parameters are the sensitive background $b_\alpha$, in units of events per mol per yr, and the sensitive exposure $\varepsilon_\alpha$, in units of mol $\cdot$ yr.
They are strongly dependent on the detector and isotope features, labeled with the index $\alpha$.
The expected number of signal and background events in the detector are then obtained as, respectively:
\begin{align}
    S_{\alpha i}(m_{\beta\beta},M_{\alpha i}) &= \ln{2} \cdot N_A \cdot \varepsilon_\alpha \cdot \left(\frac{T}{1\,{\rm yr}}\right) \cdot \widetilde\Gamma_\alpha(\mbb,M_{\alpha i}) \,,\label{eq:S}\\
    B_\alpha &= b_\alpha \cdot \varepsilon_\alpha \cdot \left(\frac{T}{1\,{\rm yr}}\right) \,,
    \label{eq:B}
\end{align}
with $N_A$ being Avogadro's number, $T$ the exposure time, and the index $i$ labels the different nuclear models from Tab.~\ref{tab:NME}.
While the background is independent of the nuclear model assumed, the signal strongly depends on it. The total number of events expected in each detector is then given by 
\begin{equation}
\label{eq:tot_ev}
N_{\alpha i} = S_{\alpha i}  + B_{\alpha}\,.
\end{equation}
Tab.~\ref{tab:experiments} summarises the performance parameters of each selected experiment and the PSF of the related isotope \cite{Deppisch:2020ztt}.
We chose next-generation experiments for our analysis by taking, for each isotope, the one with the most ambitious configuration from the ones discussed in \cite{Agostini:2022zub}. These should be considered as examples for ``ultimate'' proposals for each isotope. Our results of course also apply to other proposals, if they reach equivalent performance parameters, for instance NEXT~\cite{NEXT:2020amj}, DARWIN~\cite{DARWIN:2020jme} for \xe\ or AMoRE~\cite{Kim:2022uce} for \mo.

\begin{table}[t]
\centering
\caption{Performance parameters assumed here for different next-generation \onbb experiments \cite{Agostini:2022zub} and PSF of the respective isotope \cite{Deppisch:2020ztt}. 
}
\label{tab:experiments}
\begin{tabular}{ccccc}
\toprule
\multirow{2}*{Experiment} & \multirow{2}*{Isotope} & $\varepsilon$ & $b$ & PSF \\
                          & & [mol$\cdot$yr] & [events/(mol$\cdot$y)] & [yr$^{-1}$ eV$^{-2}$] \\
\midrule
LEGEND-1000 & \ger & $8736$ & $4.9\cdot10^{-6}$ & $2.36\cdot10^{-26}$ \\
SuperNEMO & \se & $185$ & $5.4\cdot10^{-3}$ & $10.19\cdot10^{-26}$\\
CUPID & \mo & $1717$ & $2.3\cdot10^{-4}$ & $15.91\cdot10^{-26}$\\
SNO+II & \te &$8521$ & $5.7\cdot10^{-3}$ & $14.2\cdot10^{-26}$  \\
nEXO & \xe & $13700$ & $4.0\cdot10^{-5}$ & $14.56\cdot10^{-26}$ \\
\bottomrule 
\end{tabular}
\label{tab:future-exps}
\end{table} 
To get a feeling for the expected number of events for the \onbb experiments listed in Tab.~\ref{tab:future-exps}, we provide here the event numbers for 1~yr exposure time and a reference value $(q^2 m_{\beta\beta})^{\rm True} = 40~$\meV within the IMO band: 
\begin{align}
N_{{\rm LEGEND-1000}} &= \left\{0.97 \times \left[\frac{(q^2 m_{\beta\beta})^{\rm True}}{40 ~{\text{meV}}}\right]^2 \left(\frac{M_{\text{Ge}}^{\text{long}}}{2.66}\right)^2 + 0.04  \right\}\times \frac{T}{1~{\rm yr}} \;,\label{eq:event_Ge}\\
N_{{\rm SuperNEMO}} &= \left\{0.09 \times \left[\frac{(q^2 m_{\beta\beta})^{\rm True}}{40 ~{\text{meV}}}\right]^2 \left(\frac{M_{\text{Se}}^{\text{long}}}{2.72}\right)^2 + 1.0 \right\}\times \frac{T}{1~{\rm yr}} \;,\label{eq:event_Se}\\
N_{{\rm CUPID}} &= \left\{0.92\times \left[\frac{(q^2 m_{\beta\beta})^{\rm True}}{40 ~{\text{meV}}}\right]^2 \left(\frac{M_{\text{Mo}}^{\text{long}}}{2.24}\right)^2 + 0.4 \right\}\times \frac{T}{1~{\rm yr}} \;,\label{eq:event_Mo}\\
N_{{\rm SNO+II}} &= \left\{1.51 \times \left[\frac{(q^2 m_{\beta\beta})^{\rm True}}{40 ~{\text{meV}}}\right]^2 \left(\frac{M_{\text{Te}}^{\text{long}}}{1.37}\right)^2 + 48.6 \right\}\times \frac{T}{1~{\rm yr}}\;, \label{eq:event_Te} \\
N_{{\rm nEXO}} &= \left\{1.64 \times \left[\frac{(q^2 m_{\beta\beta})^{\rm True}}{40 ~{\text{meV}}}\right]^2 \left(\frac{M_{\text{Xe}}^{\text{long}}}{1.11}\right)^2 + 0.5 \right\}\times \frac{T}{1~{\rm yr}} \label{eq:event_Xe} \;,
\end{align}
where the smallest $M_{\alpha}^{ \text{long}}$ values from Tab.~\ref{tab:NME} have been taken as normalisation.
We observe that $(q^2\mbb)^{\rm True}$ values somewhat larger than $40~$\meV and/or exposure times of several years will be required to obtain sizeable event numbers for all these five experiments. Note also the comparable large background in SNO+II, which consists of an important limitation for this project.

In order to study the sensitivity of future projects we consider the following  $\chi^2$ function, based on the Poisson distribution as required for the small number of expected events:
\begin{equation}
\label{eq:chi2}
\Delta\chi^2_{ij}(m_{\beta\beta}, M_{\alpha j} \,;\, m_{\beta\beta}^\text{True}, M_{\alpha i}^\text{True}) = 2\sum_\alpha 
\left(N_{\alpha j}- N_{\alpha i}^{\text{True}} + N_{\alpha i}^{\text{True}} \ln \frac{N_{\alpha i}^{\text{True}}}{N_{\alpha j}}
\right) \;,
\end{equation}
where $N_{\alpha i}^{\text{True}} = B_{\alpha i} + S_{\alpha i} (m_{\beta\beta}^{\rm True},M_{\alpha i}^\text{True})$ refers to the true model assumed to be realised in Nature and $N_{\alpha j} = B_{\alpha j}+ S_{\alpha j} (m_{\beta\beta},M_{\alpha j})$ denotes the model to be compared with ``data'' corresponding to the ``true'' model. Both the indices $i,j$ run over all the nuclear models considered in Tab.~\ref{tab:NME}, namely, from N1 to I2, while the sum over $\alpha$ is over the experiments (or isotopes) considered together. In our analyses we do not consider statistical fluctuations around the mean value $N_{\alpha i}^{\text{True}}$ but assume the so-called Asimov data set; therefore our sensitivity calculations correspond to the \emph{mean} expected sensitivities.

\subsection{Sensitivity to a \onbb\ signal}
\label{sec:sensitivity}

\begin{figure}[t]
	\centering
	\includegraphics[width=0.7\textwidth]{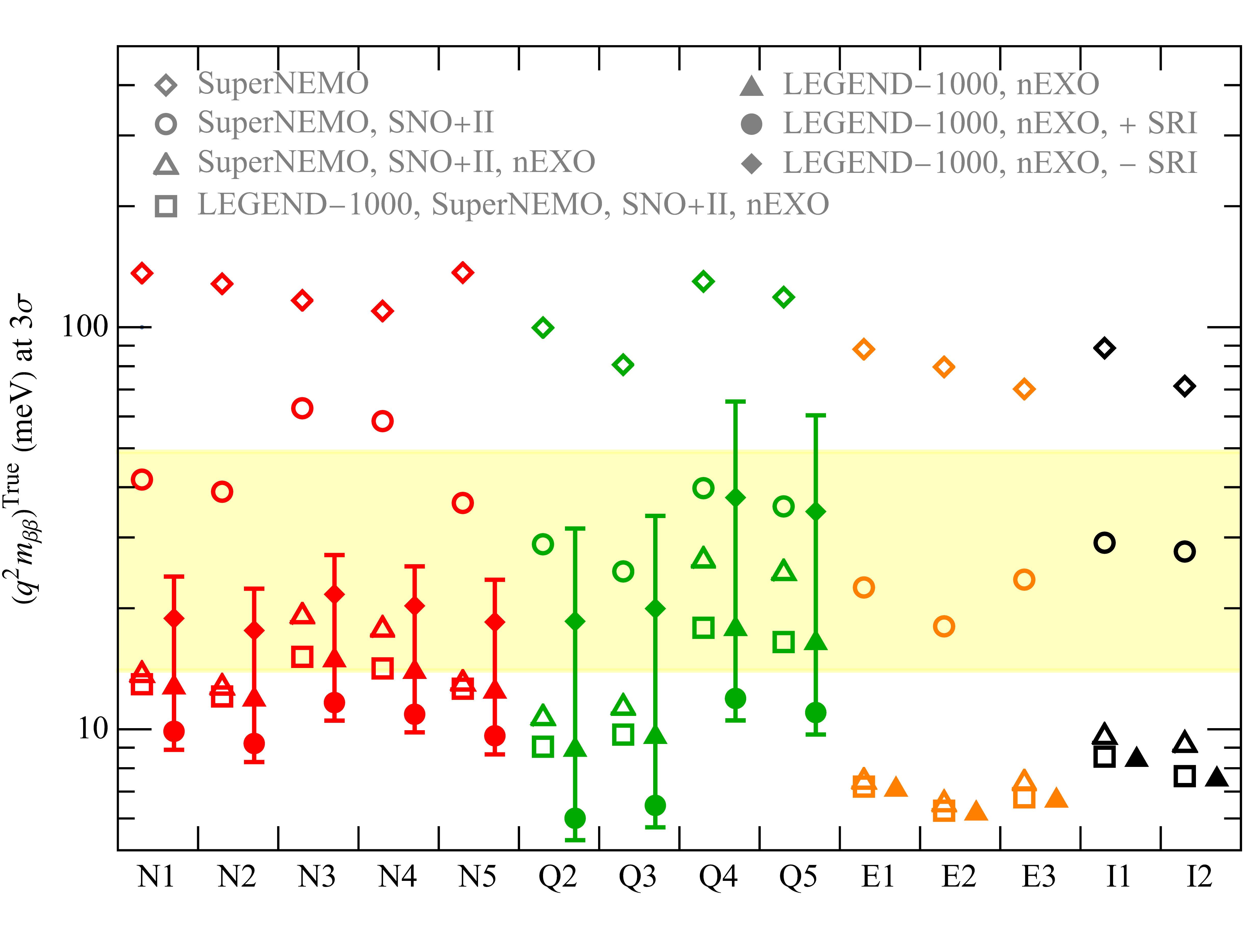}
	\caption{
 $3\sigma$ sensitivity on $(q^2 m_{\beta\beta})^{\rm True}$ for various NME models, obtained by assuming $m_{\beta\beta}=0$, $T=10~$yr and $\Delta \chi_{ij}^2 =9$ in Eq.~\eqref{eq:chi2} for different combinations of future experiments. Both cases of neglecting and including contribution from SRI are included.
 Solid circle and diamond markers represent the central value of $|n_{\alpha i}|$, while the two ends of vertical bars, obtained by assuming the boundary values of $|n_{\alpha i}|$ in the range of Tab.~\ref{tab:n_ranges} and combining LEGEND-1000 and nEXO, highlight the uncertainty on the short-range contribution. The coloured area indicates the IMO region for $q^2=1$. 
        }
 \label{fig:sensitivity_exp}
\end{figure}

\begin{figure}[t]
	\centering
	\includegraphics[width=0.98\textwidth]{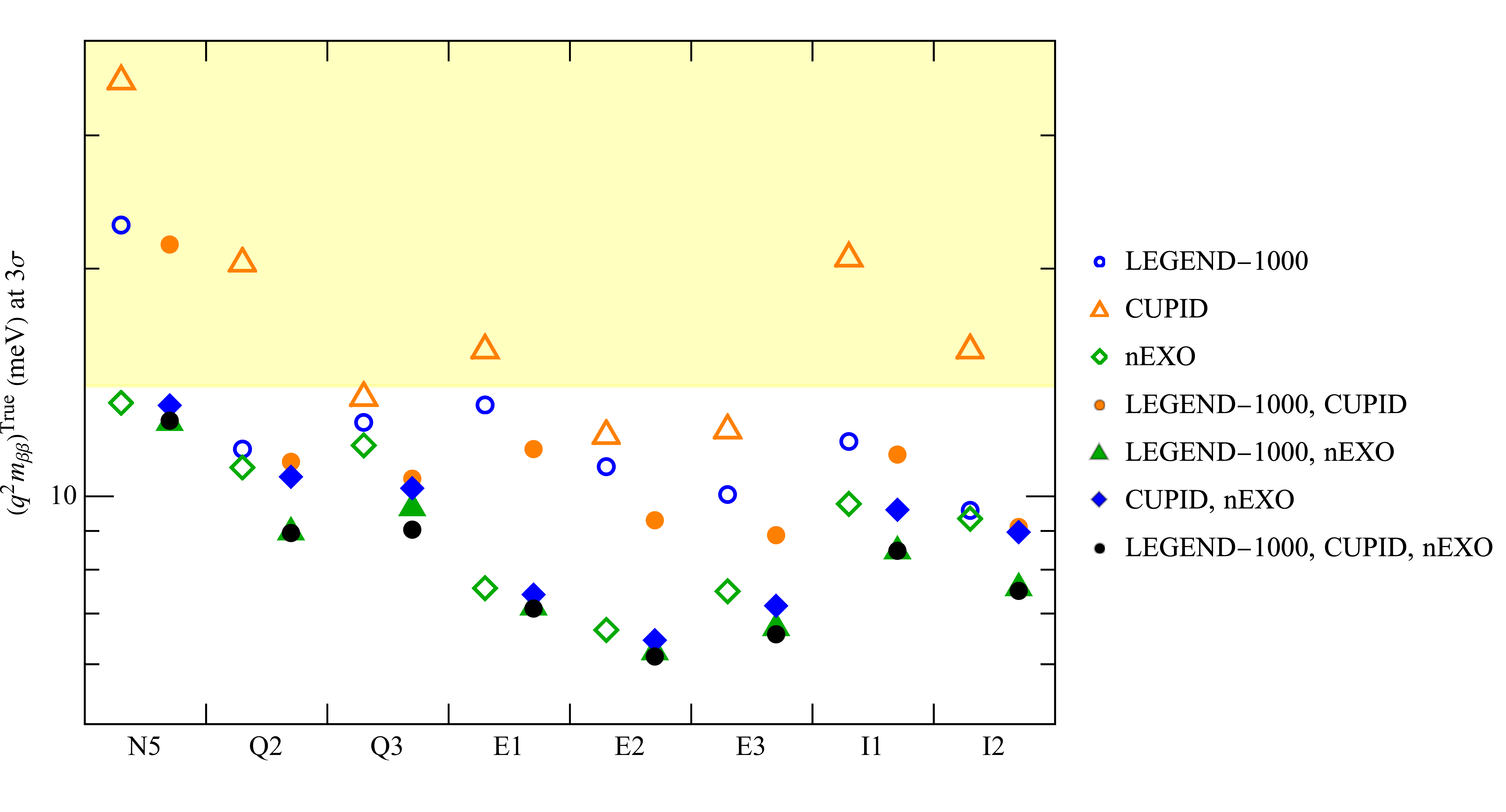}
	\caption{$3\sigma$ sensitivity for $(q^2 m_{\beta\beta})^{\rm True}$, obtained by assuming $m_{\beta\beta}=0$, $T=10~$yr and $\Delta \chi_{ij}^2 =9$ in Eq.~\eqref{eq:chi2}, by including the CUPID contribution, both in comparison and in combination with the dominant sensitivities of LEGEND-1000 and nEXO. 
 A reduced sample of nuclear models has been considered, see Tab.~\ref{tab:NME}, and contributions from SRI are neglected. The coloured area refers to the IMO region
 for $q^2=1$.        }
 \label{fig:sensitivity_exp_CUPID}
\end{figure}

Let us first study the sensitivity to a \onbb\ signal of the
considered experiments. 
To this aim, we set $\mbb = 0$ in
Eq.~\eqref{eq:chi2} and study $\Delta\chi^2_{ij}$ as a function of
$(q^2 m_{\beta\beta})^\text{True}$ and the true NME $M_{\alpha
  i}^\text{True}$. 
By imposing $\mbb=0$, $\Delta\chi^2_{ij}$ becomes
independent of the index $j$ labeling the nuclear model used for the fit. 
Hence, for a given true NME $M_{\alpha i}^\text{True}$, requiring for instance $\Delta\chi^2_{ij} \ge 9\, (25)$ will lead to the region of
$(q^2 m_{\beta\beta})^\text{True}$, for which a positive \onbb\ signal can
be established at $3\sigma$ ($5\sigma$). 
An important remark is that, as reported in Tab.~\ref{tab:NME}, some NME calculations are missing for some isotopes used by future experiments we are considering.
In particular, concerning the CUPID case, there are only 7 long-range NMEs available with which we can perform the analysis in Tab.~\ref{tab:NME}.
Therefore, to show clearly how the sensitivities change as a function of different nuclear models and combinations of experiments, we studied two cases separately, first without the CUPID contribution, and then adding it for a reduced sample of nuclear models. The corresponding results are shown in Fig.~\ref{fig:sensitivity_exp} and Fig.~\ref{fig:sensitivity_exp_CUPID}, respectively. 

Fig.~\ref{fig:sensitivity_exp} depicts the $3\sigma$ sensitivity on $(q^2 \mbb)^{\text{True}}$ for different combinations of LEGEND-1000, SNO+II, nEXO and SuperNEMO, obtained by assuming $\Delta\chi^2_{ij} = 9$, $\mbb = 0$ and an exposure time $T=10~$yr for each experiment.
Both the cases of neglecting and including the effects coming from SRI have been studied.
It results that, two or more experiments combined, as well as the inclusion of the SRI effects, allow to partially or completely cross the IMO region for most NME models. Together with Fig. \ref{fig:sensitivity_exp_CUPID},
we find the best sensitivity is given, for all the nuclear models, by the combination of LEGEND-1000 and nEXO (filled markers) if we only input two experiments of the five listed in Tab.~\ref{tab:future-exps}.
The inclusion of SuperNEMO and SNO+II does not improve significantly the sensitivity (see the comparison between the open squares and solid triangles). Again, we see the big uncertainties on the sensitivity to $m_{\beta\beta}^{\rm True}$ induced by the presence of the SRI, and especially the sign of its contribution.
Fig.~\ref{fig:sensitivity_exp_CUPID} also specifies the contribution of CUPID to the sensitivity on $(q^2\mbb)^{\rm True}$ for the restricted sample of nuclear models in Tab.~\ref{tab:NME} for which $^{100}$Mo calculations are available.
Only the LEGEND-1000 and nEXO contributions are reported here, since these emerged to be the dominant ones. Comparing the three experiments individually (open symbols), we see that CUPID has a similar sensitivity to the assumed LEGEND-1000 and nEXO configurations, though slightly worse. By comparing the $(q^2\mbb)^{\rm True}$ sensitivity from the LEGEND-1000 and nEXO combination (solid green triangles) with the combined LEGEND-1000, nEXO and CUPID one (solid black dots), the contribution of CUPID to the sensitivity will not be dominant. For some NME models (E1, E2, E3) the nEXO-CUPID combination is very powerful and adding LEGEND-1000 in addition leads to little improvement.
Note that in Fig.~\ref{fig:sensitivity_exp_CUPID} the SRI contribution is ignored for illustration purpose, because we want to specifically highlight the sensitivities of different combinations of experiments.

\begin{figure}[h]
	\centering
	\includegraphics[width=\textwidth]{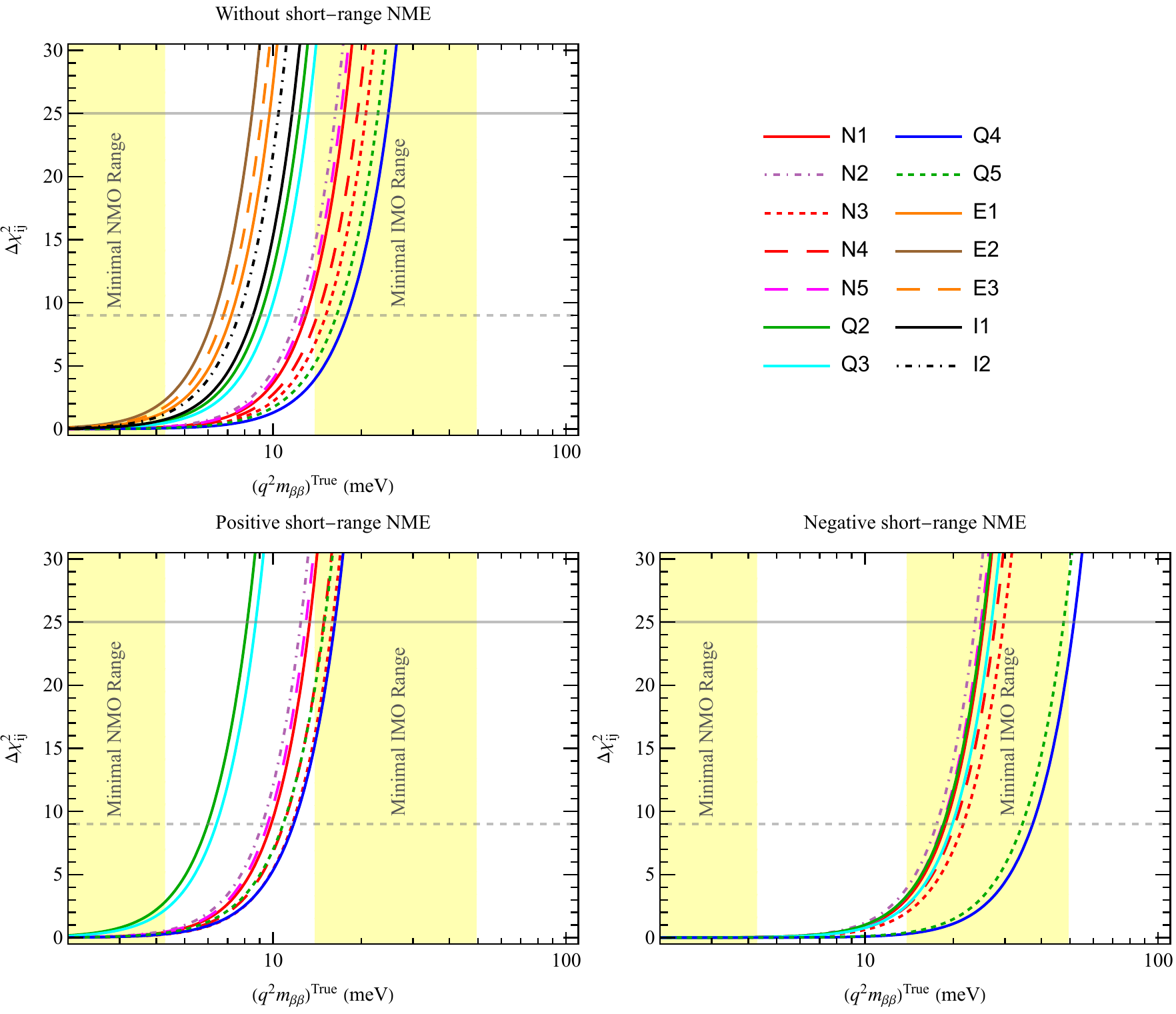}
	\caption{Values of $\Delta\chi_{ij}^2$ as a function of $(q^2 m_{\beta\beta})^{\rm True}$, taking $m_{\beta\beta}=0$ and $T=10~$yr, for the combination of LEGEND-1000, SNO+II, nEXO and SuperNEMO experiments.  
 The bottom left/right panel assumes, respectively, a positive/negative sign of $n_{\alpha i}$,while in the upper one the contribution of SRI is ignored. 
 The horizontal lines denote $3\sigma$ (dashed gray) and $5\sigma$ (solid gray) C.L.\ and the yellow areas indicate the IMO and NMO bands for $q^2=1$.
 }
	\label{fig:sensitivity_line}
\end{figure}

Complementary to Fig.~\ref{fig:sensitivity_exp}, Fig.~\ref{fig:sensitivity_line} shows how the $\Delta\chi^2$ relevant for the sensitivity changes with $(q^2 m_{\beta\beta})^{\rm True}$.  
We see that a positive contribution from SRI leads for most nuclear models to favor a \onbb signal even at $5\sigma$ in the IMO case, while a negative SRI contribution only implies the \onbb signal at $3 \sigma$ mostly. Supposing that the lightest neutrino mass is zero in the case of normal neutrino mass ordering (NMO),  we get the minimal NMO range of $m_{\beta\beta}$:
$ 0.9~{\rm meV}<m_{\beta\beta}<4.2~{\rm meV}$, similar to Eq.~\eqref{eq:mbb-IMO}. As indicated in the figure, 
it will be difficult to observe a positive \onbb signal from the considered combination of future LEGEND-1000, SNO+II, nEXO and SuperNEMO project if Nature takes values of $(q^2 m_{\beta\beta})^{\rm True}$ within the minimal NMO range.

%
\section{Discrimination of NME models}
\label{sec:NMEdiscr}

In this section we are going to address the following question. Assuming that future \onbb\ experiments detect a positive signal, will it be possible via the combination of several experiments using different isotopes to discriminate among the various NME models? As the NME for different isotopes in different nuclear models are not just proportional to each other, in principle the combination of several isotopes may allow to disfavour certain models. In the following we study this possibility quantitatively, under the assumptions of the background and exposures specified in Tab.~\ref{tab:future-exps}.
To this aim we assume a true NME model $M_{\alpha i}^\text{True}$ and a true value $(q^2 \mbb)^{\rm True}$. Then, in order to test whether data allows to disfavour a certain alternative NME model $j$, we consider the test statistic
\begin{equation}\label{eq:Dchi2min}
  (\Delta\chi^2_{ij})_{\rm min} = \min_{\mbb}
\Delta\chi^2_{ij}(m_{\beta\beta}, M_{\alpha j} \,;\, (q^2 m_{\beta\beta})^\text{True}, M_{\alpha i}^\text{True})\,.
\end{equation}
If this quantity is significantly different from zero, we will be able to exclude the NME model $j$, irrespective of possible values of $\mbb$. 
Since we minimize with respect to $\mbb$, and signal event numbers depend on the product $q^2 \mbb$, our results will be independent on the uncertainty of the quenching factor, although $q$ will impact the simulated data for an assumed value of $\mbb^{\rm True}$. 
In what follows, we assume an exposure time $T=10$~yr for all the experiments and combine all the ones for which NME calculations are available for a given choice of $(i,j)$.

\subsection{Without the short-range contribution}
\label{sec:no_short}

\begin{figure}[t]
	\centering
	\includegraphics[width=0.98\linewidth]{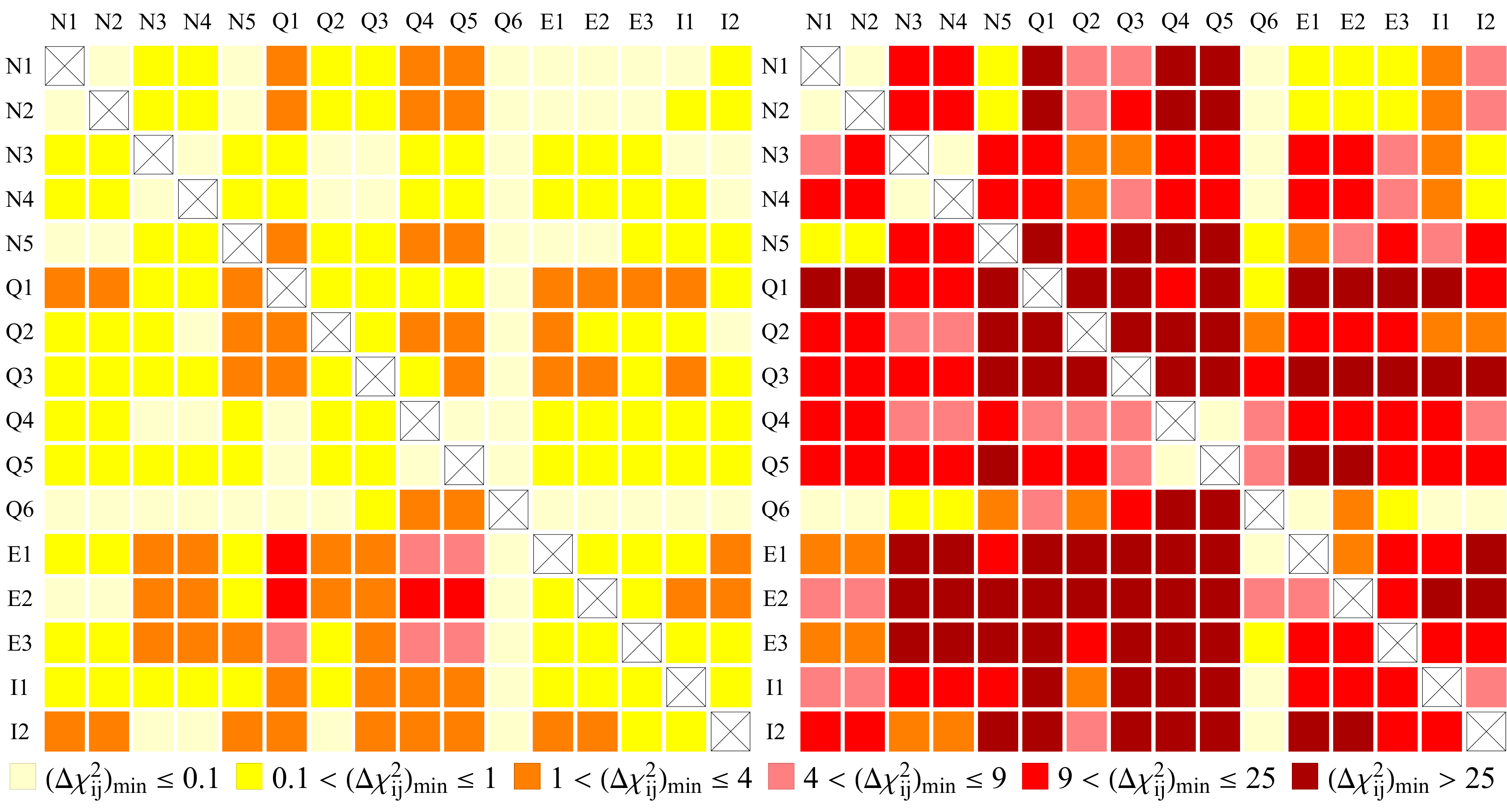}
	\caption{NME discrimination for $(q^2 m_{\beta\beta})^\text{True}  = 10~$\meV (left) and $(q^2 m_{\beta\beta})^\text{True}  = 40~$\meV (right) by taking an exposure of $T = 10~$yr.
 Indices $i$ (true model) and $j$ (fitted model) run over the vertical and horizontal axes, respectively.
 The color code shows $(\Delta \chi^2_{ij})_{\rm min}$ as defined in Eq.~\eqref{eq:Dchi2min}. 
    For each NME model combination we combined the experiments listed in Tab.~\ref{tab:future-exps} for which the corresponding NME calculations are available in Tab.~\ref{tab:NME} and neglected the short-range contribution to the NME.}
\label{fig:NMEdiscrLR}
\end{figure}

We show first results by ignoring the contribution from the short-range term to discuss in some detail the benefit of combining different elements; the impact of the SRI is highlighted in the following subsection. Results of this analysis are summarised in Fig.~\ref{fig:NMEdiscrLR} for two different assumptions on the true value of the effective Majorana mass: $ (q^2m_{\beta\beta})^{\text{True}} = 10~$\meV (left panel) and $ (q^2m_{\beta\beta})^{\rm {True}} = 40~$\meV (right panel), close to the lower and upper edges of the minimal IMO range, respectively. Coloured boxes show the discrimination potential of each model combination: reference values of $1\sigma$, $2\sigma$, $3\sigma$ and $5\sigma$ C.L., corresponding to $(\Delta\chi^2_{ij})_\text{min} =1, 4, 9$ and $25$, are shown. 
Note that, due to Poisson statistics, in general $(\Delta \chi^2_{ij})_{\text{min}} \neq (\Delta \chi^2_{ji})_{\text{min}}$. Looking at the left panel of Fig.~\ref{fig:NMEdiscrLR} with $(q^2 m_{\beta\beta})^{\rm True} = 10 ~{\rm meV}$, $(\Delta\chi_{ij}^2)_{\text{min}} \geq 9$ appears in very few model combinations. 
In general, it will be hard to rule out NME models if Nature has chosen such small values for $(q^2\mbb)^{\rm True}$. 
It becomes more promising when higher $(q^2 m_{\beta\beta})^{\text{True}}$ values are assumed, as shown in the right panel: in this case it is possible to discriminate at $3\sigma$~C.L.\ or higher for a broad range of combinations of nuclear models.

\begin{figure}[t]
	\centering
	\includegraphics[width=0.75\textwidth]{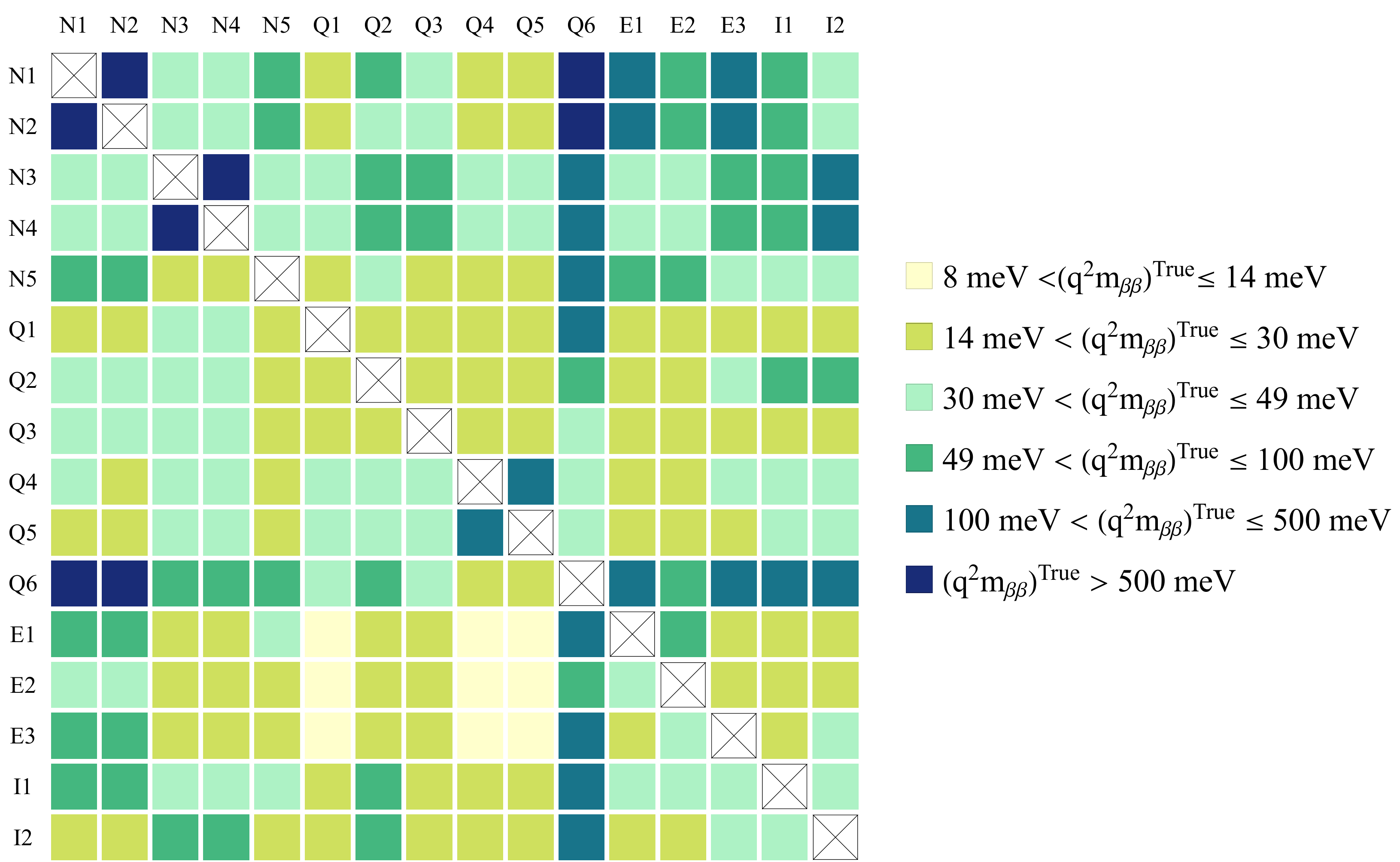}
	\caption{Values of $ (q^2 m_{\beta\beta})^{\text{True}} $ that allow nuclear model discrimination at $3\sigma$ C.L., corresponding to $(\Delta \chi_{ij}^2)_{\rm min}=9$, assuming $T = 10~$yr exposure time for all the future experiments listed in Tab.~\ref{tab:future-exps}, where contribution from SRI is neglected. 
 Indices $i$ (true model) and $j$ (fitted model) run over the vertical and horizontal axes, respectively, and refer to nuclear models in Tab.~\ref{tab:NME}. 
 }
\label{fig:chi2min-3sigma}
\end{figure}

A complementary analysis is shown in Fig.~\ref{fig:chi2min-3sigma}, where we fix $(\Delta\chi^2_{ij})_{\text{min}} = 9$ and show, by the color code, the required value of $(q^2\mbb)^{\text{True}}$ for a given nuclear model combination $i,j$. 
We find that for a large set of combinations ($i,j$) discrimination at $3\sigma$ is possible for $(q^2\mbb)^{\text{True}} > 14~$\meV (the lower bound of minimal IMO range in \ref{eq:mbb-IMO}).
The only cases which allow model discrimination assuming $(q^2\mbb)^{\text{True}} \leq 14~$\meV are $i = \{\text{E1,E2,E3}\}$, $j = \{ \text{Q1,Q4,Q5}\}$, consistent with the results of Fig.~\ref{fig:NMEdiscrLR}. This implies very promising discrimination potential even beyond the minimal IMO range in the case of these combinations.

\begin{figure}[t]
    \centering
\includegraphics[width=0.98\textwidth]{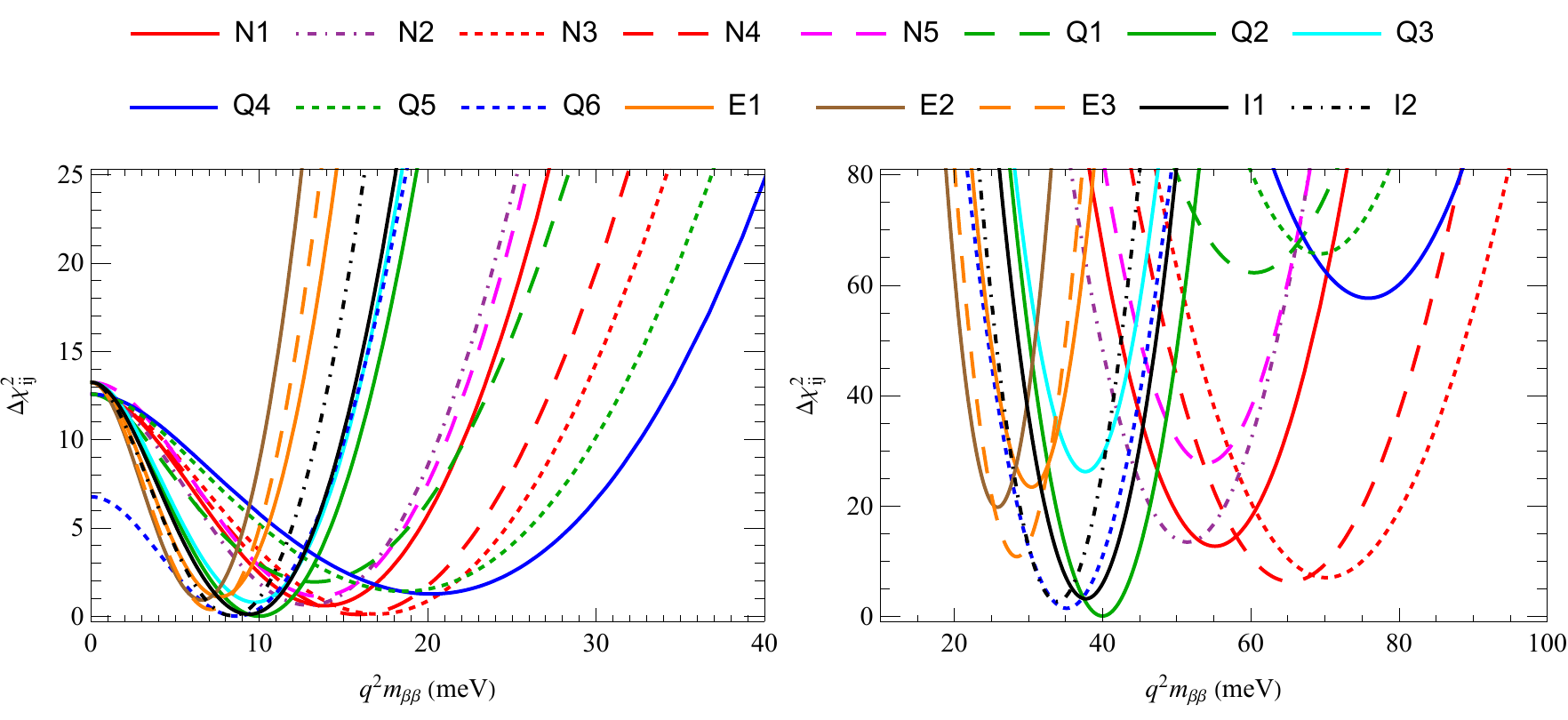}
    \caption{An example of how $\Delta \chi^2_{ij}$ change with $q^2 m_{\beta\beta}$ by assuming that the true NME model is Q2 (that is, $i=$ Q2 and $j=$ N1--I2), and taking the value of $(q^2 m_{\beta\beta})^{\rm True}$ as
    $10~$\meV (left panel) or $40~$\meV (right panel), where all the contributions from the experiments in Tab. \ref{tab:future-exps} for which the corresponding NME calculations are available in Tab. 
    \ref{tab:NME} are considered and
    the contribution of short-range NME is ignored.
    }
    \label{fig:chi2-mbb-Q2true}
\end{figure}

To see how the $\Delta\chi^2_{ij}$ in Eq.~\ref{eq:chi2} varies as a function of the fitted $q^2 \mbb $, we produced Fig.~\ref{fig:chi2-mbb-Q2true}, where we fix as an example the true model $i = \rm Q2$, and plot the $\Delta\chi^2_{ij}$ profile for each $j$ model in Tab.~\ref{tab:NME} for the two cases of $(q^2\mbb)^{\rm True}  = 10~$\meV (left panel) and $40~$\meV (right panel), always considering an exposure of $T=10~$yr and all the experiments of Tab.~\ref{tab:future-exps} in combination. 
The vertical distance between the minima of any pair of NME models corresponds to the color code in Fig.~\ref{fig:NMEdiscrLR}.
The solid green curve represents the case in which we are fitting with the ``correct'' NME model (i.e., $j=\rm Q2$); and therefore this curve recovers the minimum at $\Delta\chi^2=0$ at the assumed true value for $q^2\mbb$. For $j\neq \rm Q2$, the figure
shows how much using a ``wrong'' NME model will lead to deviations of the 
fitted $q^2 m_{\beta\beta}$ from its true value (both for cases where a statistically significant rejection of the assumed NME is or is not possible). Note that 
for the given true NME model Q2 and different fitted NME model ($j=$N1--I2), the $\Delta\chi^2_{ij}$ values at $m_{\beta\beta}=0$ do not converge to the same value. This is because we consider different combinations of the experiments depending on the availability of the corresponding NME calculation in Tab.~\ref{tab:NME}, while in Figs.~\ref{fig:sensitivity_exp}, \ref{fig:sensitivity_exp_CUPID} and \ref{fig:sensitivity_line} we always compare the sensitivities of the same combinations of proposed experiments.

\begin{figure}[t]
	\centering
	\includegraphics[width=0.98\textwidth]{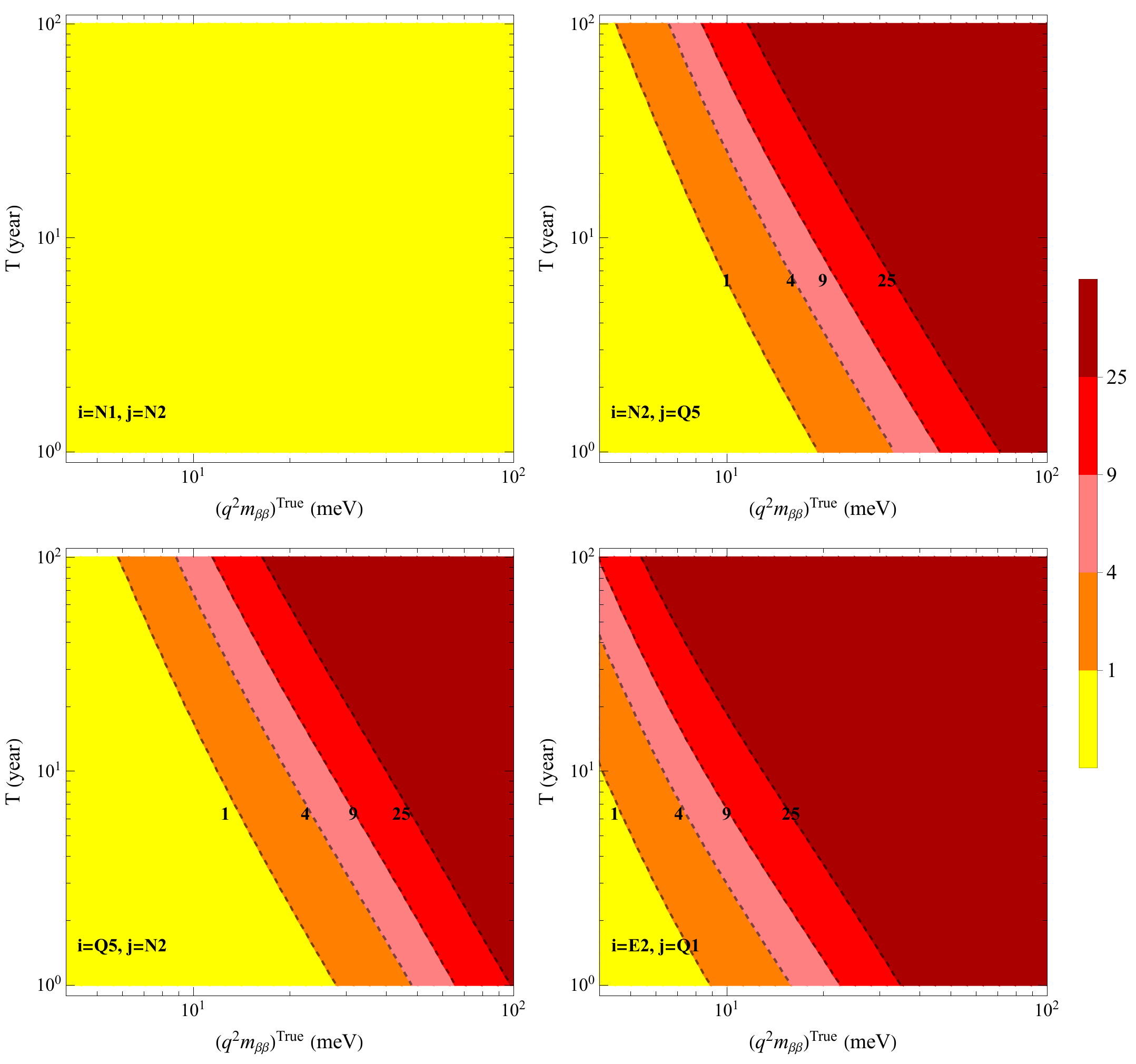}
	\caption{Contours of $(\Delta\chi_{ij}^{2})_{\text{min}}$, in the $ (q^2\mbb)^{\text{True}} - T$ plane by considering four representative nuclear model combinations $(i,j)$, in absence of SRI. For each of them, we included all possible combination of experiments allowed by data in Tab.~\ref{tab:future-exps}. 
 }
\label{fig:contours_wo_short}
\end{figure}

We emphasize that in the above analysis we have assumed ambitious experimental configurations and exposure times of 10 years. To study the dependence on these assumptions, we show the contours of $(\Delta \chi_{ij}^2)_{\rm min}$ in the plane of $(q^2 m_{\beta\beta})^{\rm True}$ and $T$, where we take the model combinations $(i,j)=$ (N1, N2), (N2, Q5), (Q5, N2) and (E2, Q1) as four typical examples. We take the exposure time $T$ to parameterize the total effective exposure, consisting as measurement time times mass of the experiment and show $T$ ranging from 1 to 100 years. In the upper left panel, we can see that it is impossible to discriminate N1 from N2 in a reasonable range of $(q^2 m_{\beta\beta})^{\rm True}$ even with a
100-year exposure time. As for the most promising case, it is likely to discriminate between E2 and Q1 at $3\sigma$ already for a 1-year exposure and $(q^2 m_{\beta\beta})^{\rm True}$ bigger than about $20~$\meV. 
Furthermore, the upper-right panel ($i=$ N2, $j=$ Q5) and
the lower-left panel ($i=$ Q5, $j=$ N2) illustrate the asymmetric effects 
when the true and fitted NME models are exchanged.

As we have seen in section~\ref{sec:sensitivity}, the sensitivity to $q^2 \mbb$ is dominated by the LEGEND-1000 and nEXO setups assumed here. This is to some extent also true for the NME discrimination power and the combination of these two experiments offers already good sensitivities. However, here the experiments with sub-leading sensitivity to $\mbb$ offer complementary information, due to the different isotopes used in SNO+II, CUPID and SuperNEMO, which enhances the discrimination power and allows to cover more combinations of true and fitted NME models, see appendix~\ref{appendix} for corresponding analyses.

\subsection{Including the short-range contribution}
\label{sec:short}
Once the short-range term is included in the analysis, the NMEs involved in calculations are defined, by Eq.~\ref{eq:M_short_long}, as the sum of both long- and short-range contributions.
In this case, data so far available and summarised in Tab.~\ref{tab:n_ranges}, reduce the analysis to a restricted class of nuclear models, namely the NSM and QRPA ones. 
Moreover, when we derive the $(\Delta \chi^2_{ij})_{\rm min}$, the lack of knowledge on the SRI contribution, which manifests in terms of both the unknown strength and sign of $n_{\alpha i}$, translates into a multi-parameter minimisation, over $\mbb{}$ and $n_{\alpha j}$, where $j$ refer to the fitted model.

\begin{figure}[t]
	\centering
	\includegraphics[width=0.9\textwidth]{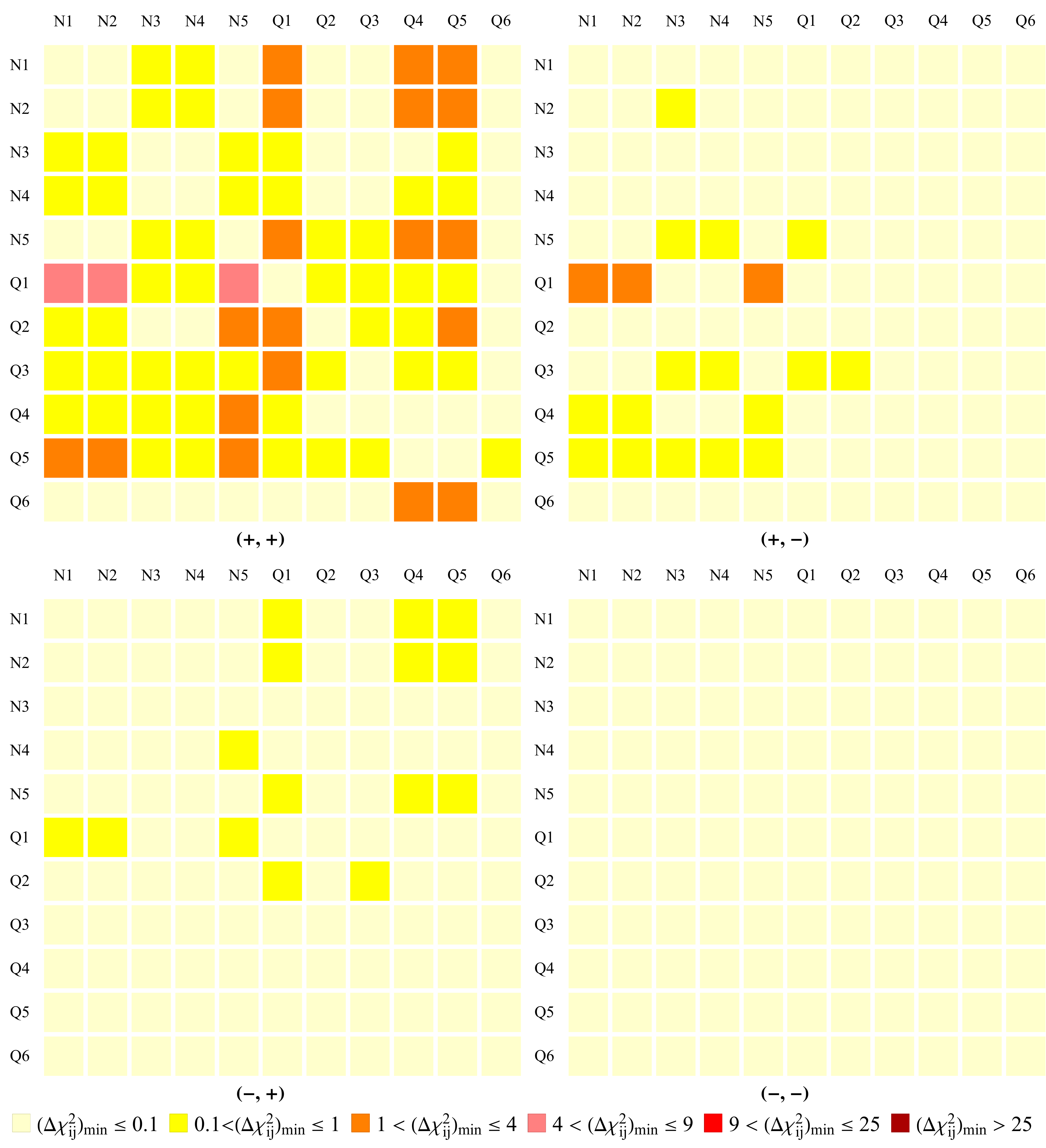}
	\caption{Minima of $\Delta \chi^2_{ij}$ for NME model discrimination when assuming $(q^2 m_{\beta\beta})^{\text{True}} = 10~$\meV, $T = 10~$yr exposure time of LEGEND-1000, SuperNEMO, CUPID, SNO+II and nEXO, 
 and including the short-range term contributions $n_{\alpha a}$, with $a = \{i,j\}$.
 Indices $i$ (true model) and $j$ (fitted model) run over the vertical and horizontal axes, respectively.
 $n_{\alpha i}^{\text{True}}$ is given by the central value of each allowed range, taken with a  positive/negative sign (the first sign in brackets), while $n_{\alpha j}$ is kept free to vary in the respective range, always with a positive/negative sign (the second sign in brackets). The four panels correspond to the combinations $(+,+)$, $(+,-)$, $(-,+)$ and $(-,-)$.
 }
\label{fig:10meV_short}
\end{figure}

\begin{figure}[t]
	\centering
	\includegraphics[width=0.9\textwidth]{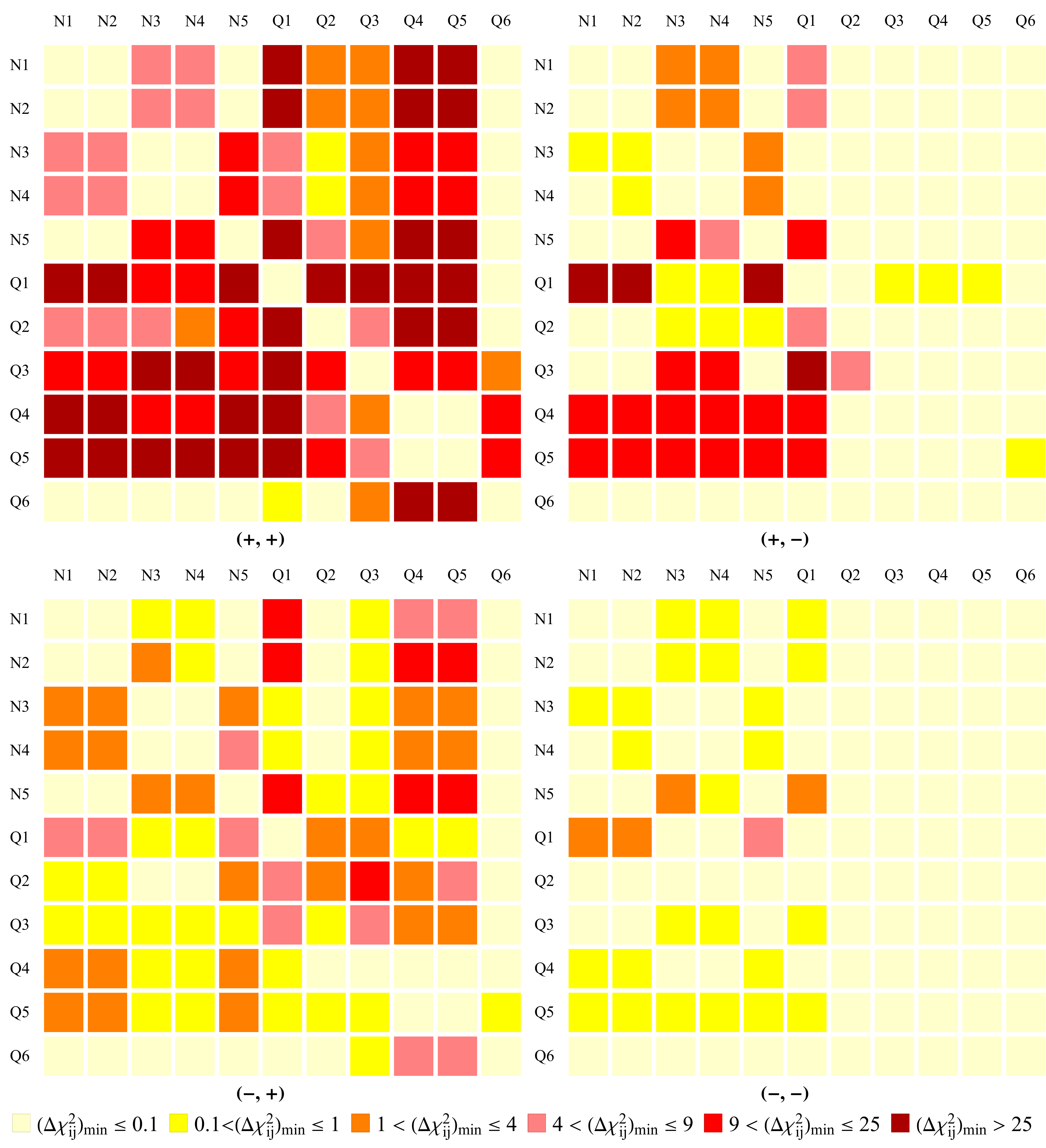}
	\caption{Minima of $\Delta \chi^2_{ij}$ when for NME model discrimination assuming $(q^2 m_{\beta\beta})^{\text{True}} = 40~$\meV, $T = 10~$yr exposure time of LEGEND-1000, SuperNEMO, CUPID, SNO+II and nEXO, and including the short-range term contributions $n_{\alpha a}$, with $a = \{i,j\}$.
 Indices $i$ (true model) and $j$ (fitted model) run over the vertical and horizontal axes, respectively.
 $n_{\alpha i}^{\text{True}}$ is given by the central value of each allowed range, taken with a  positive/negative sign (the first sign in brackets), while $n_{\alpha j}$ is kept free to vary in the respective range, always with a positive/negative sign (the second sign in brackets). The four panels correspond to the combinations $(+,+)$, $(+,-)$, $(-,+)$ and $(-,-)$.
 }
	\label{fig:40meV_short}
\end{figure}

We study the situation that the SRI contribution has the same sign for all isotopes, but the sign could be unknown.
Thus we have four cases for each value of $(q^2 m_{\beta\beta})^{\rm True}$, namely, 
$(+,+), (+ ,-)$, $(-,+)$ and $(-,-)$, where the first sign in the bracket is the assumed sign of the true SRI contribution ratio $n_{\alpha i}$, and the second one represents the sign of $n_{\alpha j}$ in the fitted NME model. 
Moreover, we take the value of $n_{\alpha i}$ (true model) at the center value of the range reported in Tab.~\ref{tab:n_ranges} and let $|n_{\alpha j}|$ for the fitted model vary in the corresponding range but keeping the same sign for all $n_{\alpha j}$. 
When comparing two NME models, we include the contribution of at most five isotopes (corresponding to the five experiments in Tab.~\ref{tab:experiments}), for each of which the long- and short- range NMEs are both available in Tabs.~\ref{tab:NME} and \ref{tab:n_ranges}. For example, when assuming that N1 is the true NME model and checking the difference with Q1, only the contributions of LEGEND-1000, SNO+II and nEXO are taken into account.
Then we study how $\Delta\chi_{ij}^2$ changes with $q^2 m_{\beta\beta}$ by fixing the model index $i$ and $(q^2 m_{\beta\beta})^{\rm True}$, similar as in Fig.~\ref{fig:chi2-mbb-Q2true}, but now each curve will correspond to a band due to the uncertainties of $|n_{\alpha j}|$. The vertical distance of the minima of any pair of NME models (two bands) determines then the corresponding $(\Delta {\chi}^2_{ij})_{\rm min}$.

Then by \st{By} performing the multi-parameter minimization of $\Delta\chi^2_{ij}$ over
$m_{\beta\beta}$ and $n_{\alpha j}$, we
show the $(\Delta\chi^2_{ij})_{\rm min}$ for different combinations of the NME models (from N1 to Q6) by the color code in Figs.~\ref{fig:10meV_short} and \ref{fig:40meV_short}, where we have
assumed $(q^2 m_{\beta\beta})^{\rm True} =10~$\meV and $40~$\meV, respectively. Compared with Fig.~\ref{fig:NMEdiscrLR}, we find that
the discrimination potential of $(+,+)$
is comparable with the case without SRI,
while $(\Delta\chi^2_{ij})_{\rm min}$ are suppressed a lot in the other three cases. The positive SRI will lead to bigger \onbb decay rate and hence more signal events while the negative one will result in smaller \onbb decay rate and damage the NME discrimination potential.
On the other hand, it is easy to see that it is not very likely to discriminate the NME models if the $(q^2 m_{\beta\beta})^{\rm True}$ smaller than $10~$\meV in the case of $(+,-)$, $(-,+)$. As for $(-,-)$, the discrimination potential is still not very promising even if we assume $(q^2 m_{\beta\beta})^{\rm True} = 40~$\meV.

\begin{figure}[t]
	\centering
\includegraphics[width=0.9\textwidth]{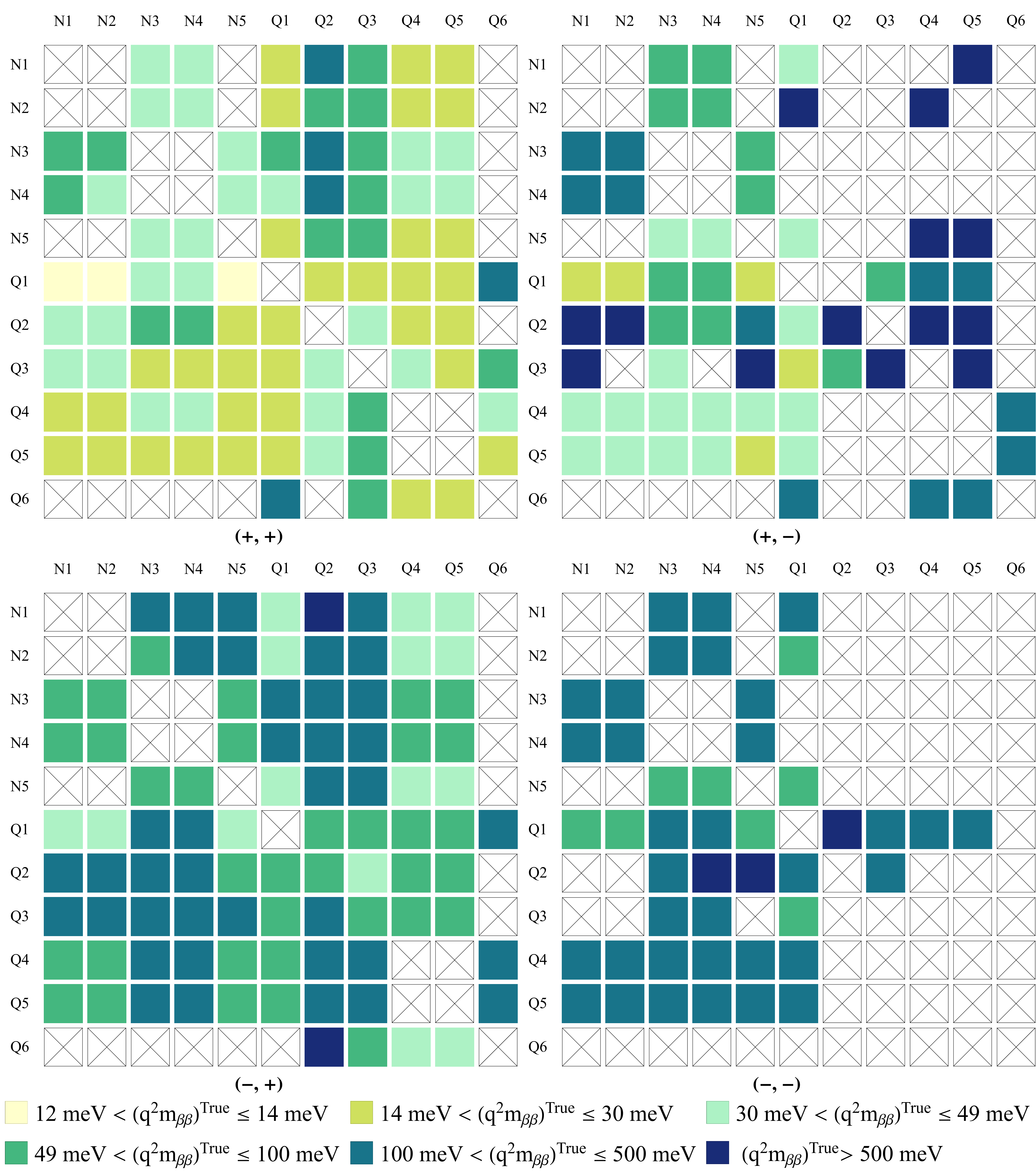}
	\caption{Values of $(q^2 m_{\beta\beta})^{\text{True}} $ that allow nuclear model discrimination at $3\sigma$ C.L., corresponding to $(\Delta \chi_{ij}^2)_{\rm min}=9$, assuming the exposure time $T = 10~$yr of LEGEND-1000, SuperNEMO, CUPID, SNO+II and nEXO, and including the short-range term contributions $n_{\alpha a}$, with $a = \{i,j\}$.
 Indices $i$ (true model) and $j$ (fitted model) run over the vertical and horizontal axes, respectively.
 $n_{\alpha i}^{\text{True}}$ is given by the central value of each allowed range, taken with a  positive/negative sign (the first sign in brackets), while $n_{\alpha j}$ is kept free to vary in the respective range, always with a positive/negative sign (the second sign in brackets). The four panels correspond to the combinations $(+,+)$, $(+,-)$, $(-,+)$ and $(-,-)$. Off-diagonal crossed squares indicate the cases where discrimination at $3\sigma$ is not possible for realistic values of $\mbb^{\rm True}$.}
  \label{fig:chi2_9_short}
\end{figure}

In Fig.~\ref{fig:chi2_9_short} we illustrate the values of $(q^2 m_{\beta\beta})^{\rm True} $ corresponding to $(\Delta\chi^2_{ij})_{\rm min}=9$ in the four cases. It suggests that it is promising to tell most NME combinations apart in the case of $(+,+)$ when $(q^2 m_{\beta\beta})^{\rm True}$ is assumed in the minimal IMO range, while discrimination for reasonable values of $(q^2 m_{\beta\beta})^{\rm True}$ is possible only for a few combinations of NME models in the cases of $(+,-)$, $(-,+)$ and $(-,-)$, consistent with Figs.~\ref{fig:10meV_short} and \ref{fig:40meV_short}.  
As for the contours of $(\Delta \chi^2_{ij})_{\rm min}$ in the $(q^2 m_{\beta\beta})^{\rm True} -T$ plane considering both long- and short NME contributions, we can get promising results similar to the $(+,+)$ case in Fig.~\ref{fig:contours_wo_short}, while the discrimination power will be severely weakened in the other three cases, as already illustrated by Figs.~\ref{fig:10meV_short}, \ref{fig:40meV_short}, and \ref{fig:chi2_9_short}.

Note that the opposite-sign cases $(-,+)$ and $(+,-)$ in Figs.~\ref{fig:10meV_short}, \ref{fig:40meV_short} and \ref{fig:chi2_9_short} show also the possibility to determine the sign of the SRI by data, as in these analyses we try to fit the simulated data with the ``wrong'' sign of the SRI. We conclude that it will be difficult to determine the sign experimentally, highlighting again the importance of theoretical and/or complementary experimental insight into this question.

\section{Conclusions}
\label{sec:conclusions}

In this work, we have studied the impact of nuclear matrix element (NME) uncertainties on the interpretation of present and future \onbb\ experiments. We compare a large set of different NME calculations (Tab.~\ref{tab:NME}) and study the corresponding variations of the bound on the effective Majorana mass $\mbb$ from present experiments (section~\ref{sec:current}) as well as the sensitivity of the next generation of experiments (section~\ref{sec:future}). In both cases we perform combined statistical analyses, combining the results or sensitivities of different experiments within a given NME model. In particular, we focus on the recently discovered short-range contribution to the NME \cite{Cirigliano:2018hja,Cirigliano:2019vdj}, which introduces an additional uncertainty in the interpretation. Finally, we investigate the possibility if in future experiments a positive \onbb signal is found, we can distinguish between different NME models by combining data from experiments using different double-beta decaying isotopes (section~\ref{sec:NMEdiscr}). 

Let us summarize our main results:
\begin{itemize}
    \item NME uncertainties due to the short-range contributions have a dramatic impact on the present combined bound on $q^2\mbb$, which may vary by a factor of order 10, depending on the sign of the short-range term. For some NME calculations the combined $3\sigma$ bound on $q^2\mbb$ reaches 40~meV, already within the upper edge of the region predicted for inverted mass ordering and vanishing lightest neutrino mass. However, from the worst case combinations the bound can become as weak as 600~meV, see Fig.~\ref{fig:mbb_upper}. 
    \item We have considered a set of advanced next generation \onbb\ experiments \cite{Agostini:2022zub} using the five isotopes $^{76}{\rm Ge}, ^{82}{\rm Se}, ^{100}{\rm Mo}, ^{130}{\rm Te}$ and $^{136}{\rm Xe}$, see Tab.~\ref{tab:future-exps} for the assumed experimental parameters. The most sensitive projects are LEGEND-1000 and nEXO, whose sensitivity to $q^2\mbb$ will cover most part of the inverted mass ordering region for many NME models, down to $q^2 \mbb \simeq 14$~meV. However, for certain NME models and unfortunate short-range interaction interference, even these advanced setups might not be able to reach the inverted mass ordering region, see Figs.~\ref{fig:sensitivity_exp}, \ref{fig:sensitivity_exp_CUPID}.
    \item Discriminating between different NME calculations will be possible for a sizeable fraction of NME model pairs by combining future experiments with different isotopes if $(q^2\mbb)^{\rm True} \gtrsim 40$~\meV. However, the presence of the short-range contribution will essentially destroy this sensitivity, unless its sign is positive and known to be positive. 
    \item Also for NME discrimination, the combination of LEGEND-1000 with nEXO is already quite powerful, but the addition of information from a third isotope, e.g., \mo\ for CUPID or \te\ for SNO+II, can significantly improve the sensitivity and increase the number of distinguishable models.
\end{itemize}

In conclusion, our work shows the crucial impact of the short-range contribution to the NMEs. In particular, as long as the sign of this contribution is not known, this introduces an uncertainty in $\mbb$ which can be as large as a factor 10 and a data-based nuclear model discrimination becomes essentially impossible. We hope that our work will stimulate further investigations towards the better understanding of the short-range NME contribution of light Majorana neutrino masses.

\section*{Acknowledgments}

We thank Dong-Liang Fang, Javier Menendez and Bernd Schwingenheuer for useful discussions and comments on the manuscript.
This work has been supported by the European Union’s Framework Programme for Research and Innovation Horizon 2020 under grant H2020-MSCA-ITN-2019/860881-HIDDeN. 
Jing-yu Zhu is supported partly by the China and Germany Postdoctoral Exchange Program from the Office of China Postdoctoral Council and the Helmholtz Centre under Grant No. 2020031 and by the National Natural Science Foundation of China under Grant No. 11835005 and 11947227.

\appendix
\section{NME model discrimination for different experiment combinations}
\label{appendix}

In this appendix, we show the nuclear model discrimination potential if different combinations of future experiments in Tab.~\ref{tab:future-exps} are considered.
According to Eq.~\eqref{eq:chi2}, the reason why we can try to tell different NME models apart through $\Delta\chi_{ij}^2$ is due to the non-linear relationship between the NMEs for different isotope. Thus we need at least two experiments using different isotopes for the discrimination. 
Here we present some results, similar to Fig.~\ref{fig:NMEdiscrLR}, but by considering a few selected pairs and triples of experiments from Tab.~\ref{tab:future-exps}. One aspect of this analysis is to study which combination of isotopes provides the best discrimination power. But of course also the assumed exposures and background levels play an important role. For simplicity, we neglect the contribution from the short-range NME, and take $(q^2 m_{\beta\beta} )^{\rm True} = 40~$\meV and an exposure of $T=10$~yr as an example. 

\begin{figure}[t]
\centering
\includegraphics[width=0.7\textwidth]{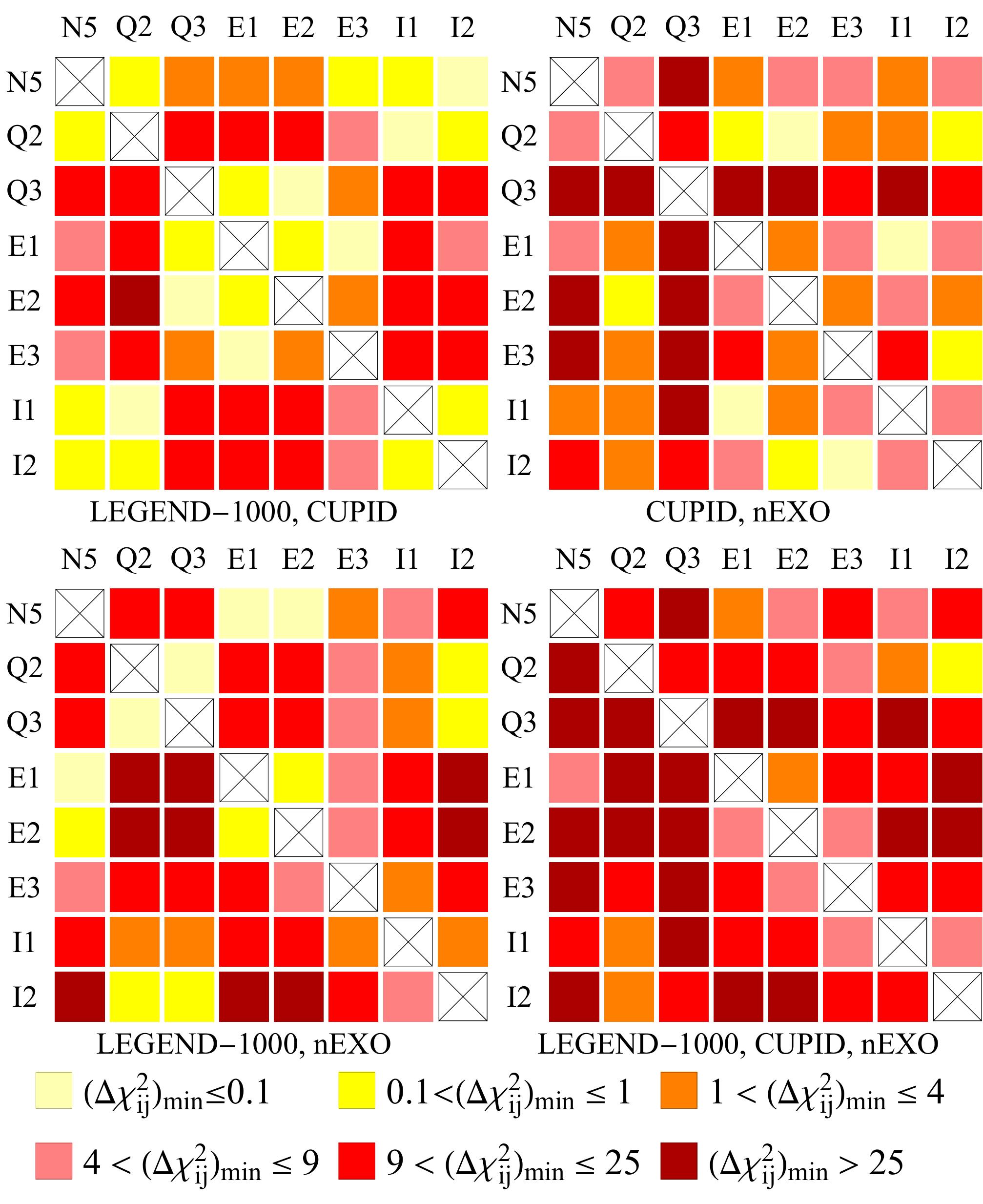}
\caption{NME discrimination for $(q^2 m_{\beta\beta})^{\rm True} = 40~$\meV by taking an exposure of $T=10$~yr and considering different combinations of LEGEND-1000, nEXO and CUPID. We show only the seven NME models for which the \mo\ NMEs relevant for CUPID are available (Q2, Q3, E1, E2, E3, I1 and I2), see Tab.~\ref{tab:NME}.} 
\label{fig:with_CUPID}
\end{figure}

\begin{figure}[t]
\centering
\includegraphics[width=0.95\textwidth]{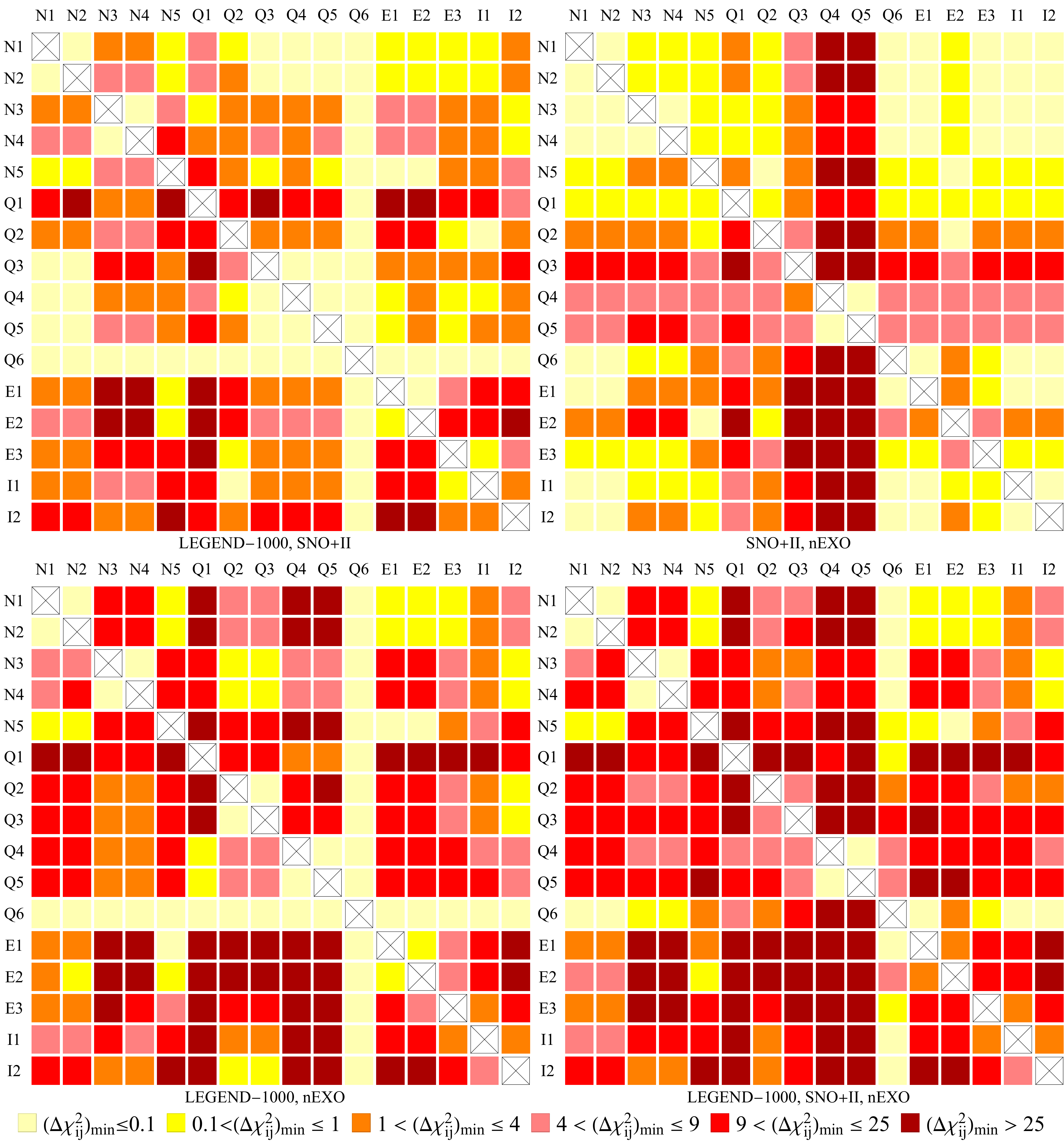}
\caption{NME discrimination for $(q^2 m_{\beta\beta})^{\rm True} = 40~$\meV by taking an exposure of $T=10$~yr and considering different combinations of LEGEND-1000, nEXO and SNO+II.}
\label{fig:SNO}
\end{figure}

As we have seen in the main part of the paper, our assumed LEGEND-1000 and nEXO setups dominate the sensitivity to $\mbb$, while the other experiments play a sub-leading role. Here we want to study the additional NME model discrimination power provided by adding the complementary information from an additional isotope to the \ger, \xe\ combination from LEGEND-1000 and nEXO, namely CUPID with \mo\ in Fig.~\ref{fig:with_CUPID}
and SNO+II with \te\ in Fig.~\ref{fig:SNO}. From comparing the 
lower left and lower right panels of these figures, we see that the involvement of the third experiment significantly improves 
the discrimination potential and allows to cover more combinations of true versus fitted NME models.

We do not show explicitly results including SuperNEMO here since with the specifications given in Tab.~\ref{tab:future-exps} we find that its contribution to nuclear model discrimination is very small. Note that the case including SRI is expected to suggest the same best combinations of experiments to discriminate NME models.

\bibliographystyle{JHEP_improved}
\bibliography{biblio}

\end{document}